\documentclass[11pt]{article}
\usepackage{fullpage}
\usepackage{cite}
\usepackage{graphicx}
\usepackage{amsmath}
\usepackage{amssymb}
\allowdisplaybreaks
\title{}
\date{}

\begin{document}
\bibliographystyle{utphys}
\newcommand{\msbar}{\ensuremath{\overline{\text{MS}}}}
\newcommand{\DIS}{\ensuremath{\text{DIS}}}
\newcommand{\abar}{\ensuremath{\bar{\alpha}_S}}
\newcommand{\bb}{\ensuremath{\bar{\beta}_0}}
\newcommand{\rc}{\ensuremath{r_{\text{cut}}}}
\newcommand{\Nd}{\ensuremath{N_{\text{d.o.f.}}}}
\setlength{\parindent}{0pt}

\titlepage

\vspace*{0.5cm}

\begin{center}
{\Large \bf BCJ duality and the double copy in the soft limit}

\vspace*{1cm}
\textsc{S. Oxburgh and C. D. White\footnote{Christopher.White@glasgow.ac.uk}} \\

\vspace*{0.5cm} SUPA, School of Physics and Astronomy, University of Glasgow,\\ Glasgow G12 8QQ, Scotland, UK\\

\end{center}

\vspace*{0.5cm}

\begin{abstract}
We examine the structure of infrared singularities in QCD and quantum General 
Relativity, from the point of view of the recently conjectured {\it double 
copy} property which relates scattering amplitudes in non-Abelian gauge 
theories with gravitational counterparts. We show that IR divergences in both 
theories are consistent with the double copy procedure, to all orders in 
perturbation theory, thus providing all loop-level evidence for the conjecture.
We further comment on the relevance, or otherwise, to the so-called 
{\it dipole formula}, a conjecture for the complete structure of IR 
singularities in QCD.
\end{abstract}

\vspace*{0.5cm}

\section{Introduction}
Scattering amplitudes in quantum field theories continue to generate
a large amount of interest, due to the many phenomenological and theoretical
applications. Much of this work (motivated by the relevance of QCD to hadron
colliders) focuses on non-Abelian gauge theories, including supersymmetric
counterparts (e.g. ${\cal N}=4$ Super-Yang Mills theory) whose simpler 
structure provides a useful testing ground for theoretical techniques. 
Amplitudes have also been studied in gravitational field theories, namely 
General Relativity and its supersymmetric counterparts such as ${\cal N}=8$
supergravity, which may provide an ultraviolet finite field theory of 
gravity~\cite{Bern:2006kd}. Although gravitational theories look superficially 
very different from non-Abelian gauge theories, a recent programme of work has 
suggested that gauge and gravity amplitudes may be related in an intriguing 
way~\cite{Bern:2010ue,Bern:2010yg,Bern:2012uf}. Firstly, gauge theory 
amplitudes may be written in a special form which manifests an explicit duality
between colour and kinematics, the so-called {\it BCJ duality} 
of~\cite{Bern:2008qj}. Secondly, an $m$-point, $L$-loop order gauge theory 
amplitude in BCJ dual form can be translated into an equivalent gravity 
amplitude by the {\it double copy} procedure, in which colour factors are
replaced by kinematic factors. That gauge theory amplitudes admit BCJ duality
is known to be the case at tree level~\cite{Bern:2010yg,Tye:2010dd,
BjerrumBohr:2010zs,Broedel:2011pd,Monteiro:2011pc}, where the double copy
property is equivalent to the known KLT relations~\cite{Kawai:1985xq} relating 
amplitudes in gauge and gravity theories. At loop level there is no formal
proof of the double copy conjecture, although it is known that this should 
hold pending the existence of BCJ duality to all orders in the gauge theory,
for the cases of pure gravity and ${\cal N}=8$ 
super-gravity~\cite{Bern:2010yg}. Loop-level checks of the duality have been 
carried out up to four loop order in ${\cal N}=4$ SYM 
theory~\cite{Bern:2010ue,Bern:1998ug,Green:1982sw,Bern:1997nh} for amplitudes involving up to five external particles, and two loop order (for four point 
scattering with maximal helicity violation) in QCD~\cite{Bern:2010ue}.\\

As is perhaps clear from the above comments, much work on BCJ duality and the 
double copy property has focused on ${\cal N}=4$ SYM theory, in which a 
restricted set of loop diagrams contribute~\cite{Bern:2012uf}. A 
significant motivation for studying ${\cal N}=4$ SYM is that the double copy
relates this to ${\cal N}=8$ supergravity, so that one may investigate the
issue of whether the latter theory is ultraviolet 
finite~\cite{Bern:2006kd,Bern:2007hh,Bern:2009kf}. To date, less work has been
invested in examining the non-supersymmetric context of pure Yang-Mills theory,
where the double copy relates this to General Relativity. Examples 
include~\cite{BoucherVeronneau:2011qv,Naculich:2011my} (see 
also~\cite{Naculich:2008ys}), 
in which amplitudes in $4\leq{\cal N}\leq 8$ supergravity are considered by 
combining BCJ dual ${\cal N}=4$ SYM amplitudes with those in a less 
supersymmetric theory, exploiting the fact that the two sets of kinematic 
factors in the double copy procedure need not both satisfy BCJ duality. The
double copy has also been investigated in the context of ${\cal N}=1$ and 
${\cal N}=2$ SYM theory~\cite{Carrasco:2012ca}. 
Recently, an extension of the duality was considered, to Yang-Mills theory
deformed by higher dimensional operators~\cite{Broedel:2012rc}.\\

The aim of this paper is to point out that the infrared limit of 
QCD~\footnote{We consider pure gluodynamics (QCD without quarks) throughout
the paper. For convenience, we refer to this as QCD throughout.} 
provides a clean environment in which to examine the double copy property in a 
non-supersymmetric context. We will argue that the known structure of infrared
singularities in both QCD and gravity is consistent with the double copy,
via BCJ duality, to all loop orders. The infrared behaviour of gauge theory 
amplitudes has been intensively studied over many years 
(see e.g.~\cite{Grammer:1973db,Mueller:1979ih,Collins:1980ih,Sen:1981sd,
Sen:1982bt,Gatheral:1983cz,Frenkel:1984pz,Magnea:1990zb,Mitov:2010rp,
Gardi:2010rn}). In QCD, the state of current knowledge regarding massless
scattering amplitudes can be summarised in the so-called 
{\it dipole formula}~\cite{Dixon:2008gr,Becher:2009cu,Becher:2009qa,
Gardi:2009qi,Gardi:2009zv}, a conjecture which states that soft and collinear
singularities exponentiate such that the exponent contains colour correlations
between at most pairs of external legs. The formula is known to be exact 
up to two loops, and the structure of possible corrections at three loop order 
and beyond has been further studied in~\cite{Dixon:2009ur,Bret:2011xm,
DelDuca:2011ae,Vernazza:2011aa,Ahrens:2012qz}. It is known that the dipole 
formula fails already at two loop level for massive external 
particles~\cite{Kidonakis:2009ev,Mitov:2009sv,Becher:2009kw,Beneke:2009rj,
Czakon:2009zw,Ferroglia:2009ep,Ferroglia:2009ii,Chiu:2009mg,Mitov:2010xw}, a 
fact which is not immediately relevant here due to the fact that the framework 
of BCJ duality and the double copy procedure is set up only for massless 
particles. \\

The study of infrared singularities in General Relativity was first examined 
in~\cite{Weinberg:1965nx}, and there has recently been a revival of interest, 
particularly in the work of~\cite{Naculich:2011ry,White:2011yy,Akhoury:2011kq}
which aims to describe soft graviton physics in the same language used in a 
gauge theory context, and which we will review in what follows. It is now well 
established that the soft limit of gravity is {\it one loop exact}. That is, 
infrared singularities exponentiate such that the exponent contains only one 
loop graphs. Note that this implies that all IR singularities in gravity are 
dipoles, in the sense that at one loop in the exponent only pairs of particles
can be correlated. \\

The above comments, and the relationship between gauge theory and gravity 
offered by the double copy procedure, beg the question: could the QCD dipole
formula have an essentially gravitational origin? This thought motivated the 
present study, and we will see that in fact the answer appears to be no. 
Nevertheless, we will find that the structure of IR singularities in QCD match
up at all orders with the known structure of singularities in General 
Relativity, providing all-loop-level evidence for the double copy conjecture.
The argument is insensitive to possible corrections to the dipole formula - we
will see explicitly that many singularities disappear when the double copy is 
taken. This includes collinear singularities, which are already known to
vanish in gravity after summing over 
diagrams~\cite{Weinberg:1965nx,Akhoury:2011kq}. \\

The structure of the paper is as follows. In section~\ref{sec:review} we 
review necessary background material in more detail, namely concepts relating 
to BCJ duality, the double copy, and the structure of infrared singularities 
in QCD and gravity. In section~\ref{sec:1loop} we examine BCJ duality and the
double copy in the soft limit at one loop order, before extending this analysis
to two loop order in section~\ref{sec:2loop}. In section~\ref{sec:gen} we 
generalise our remarks to all loop orders, before summarising our results and
concluding in section~\ref{sec:conclude}.

\section{Review of necessary concepts}
\label{sec:review}
\subsection{BCJ duality and the double copy}
\label{sec:BCJ}
In this section we briefly summarise BCJ duality~\cite{Bern:2008qj} and 
the associated double copy conjecture~\cite{Bern:2010ue,Bern:2010yg}, which
together provide a map from gauge theory to gravity amplitudes, potentially at
the multiloop level~\footnote{See also section II(A) of~\cite{Bern:2012uf} for 
a recent and pedagogical review of this material.}. \\

\begin{figure}
\begin{center}
\scalebox{1.0}{\includegraphics{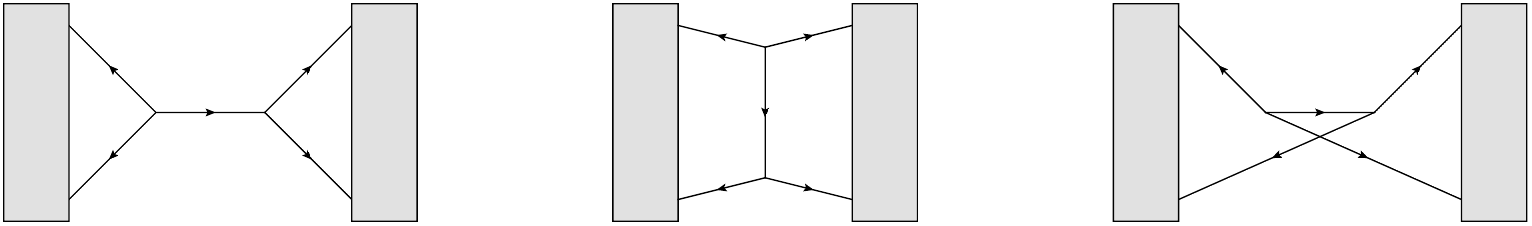}}
\caption{Illustration of three graphs related by BCJ transformations,
where the grey boxes denote the rest of the amplitude.}
\label{stufig}
\end{center}
\end{figure}

A general massless $m$-point $L$-loop gauge theory amplitude ${\cal A}_m^{(L)}$
in $D$ space-time dimensions can be written as
\begin{equation}
{\cal A}_m^{(L)}=i^Lg^{m-2+2L}\sum_{i\in\Gamma}\int\prod_{l=1}^L\frac{d^Dp_l}
{(2\pi)^D}\frac{1}{S_i}\frac{n_i\,C_i}{\prod_{\alpha_i}p_{\alpha_i}^2},
\label{ampform}
\end{equation}
where $g$ is the coupling constant. Here the sum is over the complete set
of graphs involving triple gluon vertices, consistent with the given loop 
order and number of external particles, and $S_i$ a symmetry factor for each
graph $i$ (the dimension of its automorphism group). The denominator contains
all relevant propagator momenta, and $n_i$ is the kinematic numerator 
associated with each graph. Finally, $C_i$ is the colour factor of each graph,
obtained by dressing each three gluon vertex with a factor (adopting the same 
conventions as~\cite{Bern:2012uf})
\begin{equation}
\tilde{f}^{abc}=i\sqrt{2} f^{abc},
\label{colfac}
\end{equation}
where $f^{abc}$ are the usual SU(3) structure constants. Noteworthy points
regarding this formula are:
\begin{itemize}
\item The restriction to graphs with cubic vertices only does not mean that
graphs with four gluon vertices have been excluded. Rather, one can always 
choose to rewrite the latter in terms of cubic graphs, by introducing extra
momenta in the denominator (which are compensated by additional factors in the
numerator). This relies upon the fact that the colour factor for a four gluon
vertex is a product of two structure constants, consistent with a pair of
three gluon vertices.
\item The numerators in eq.~(\ref{ampform}) may or may not come from individual
Feynman diagrams. They may also have been obtained from e.g. a generalised 
unitary-based approach (see e.g.~\cite{Bern:2011qt} and references therein), 
in which case each $n_i$ collects all kinematic 
information associated with a particular scalar integral (fixed by the 
denominator). 
\end{itemize}
From any graph at a given loop order, one may construct two other graphs
related by taking {\it BCJ transformations} involving crossing of $s$-channel
like subgraphs into $t$ and $u$-channel subgraphs. This is best illustrated
pictorially, as in figure~\ref{stufig}.
Such transformations exist for any internal line (propagator) in a given cubic
graph, leading to a set of inter-related graphs, which can be generated from
a single diagram. The colour factors associated with the three graphs in 
figure~\ref{stufig} satisfy the identity
\begin{equation}
c_s=c_t+c_u,
\label{jacobi}
\end{equation}
where $c_i$ is the colour factor associated with the diagram containing an 
$i$-channel-like subgraph. This follows from the Jacobi identity for the 
structure constants, after factoring out the part of each colour factor which 
is the same for each graph. Note there is an ambiguity in how one defines 
signs in this identity (which can be compensated by a corresponding choice for
the numerators).  \\

It is conjectured that one can always choose to redefine the numerators $n_i$
in the amplitude of eq.~(\ref{ampform}), such that they satisfy a similar
identity to eq.~(\ref{jacobi}). That is, the numerators of the $s$, $t$ and
$u$-channel-like graphs in figure~\ref{stufig} can be chosen to obey
\begin{equation}
n_s=n_t+n_u.
\label{jacobi2}
\end{equation}
Furthermore, if a given colour factor picks up a minus sign under reordering
of a vertex, then
\begin{equation}
c_i\rightarrow -c_i\quad\Rightarrow\quad n_i\rightarrow-n_i.
\label{niflip}
\end{equation}
These properties are collectively known as {\it BCJ duality}, 
after~\cite{Bern:2008qj}. It has been proven at tree-level, but remains a 
conjecture at loop level. Transformations required to write the numerators in 
a suitable form are known as {\it generalised gauge transformations}, and an 
algorithmic procedure exists in principle to establish, for a given loop-level 
amplitude, what a possible set of numerators actually is~\cite{Bern:2012uf}. It
is not clear, however, whether this algorithm is fully general, particularly 
in non-supersymmetric theories where more diagrams enter than in supersymmetric
cases. In general, one may represent the effect of a generalised gauge 
transformation on a given numerator $n_i$ as  
\begin{equation}
n_i\rightarrow n_i+\Delta_i,
\label{gengauge}
\end{equation}
where the $\{\Delta_i\}$ must satisfy
\begin{equation}
\sum_{i\in\Gamma}\frac{\Delta_i\,c_i}{\prod_{\alpha_i}p_{\alpha_i}^2}=0.
\label{delsum0}
\end{equation}
Again the sum is over all cubic diagrams with colour factors $c_i$ and 
propagator denominators $p_{\alpha_i}^2$. Indeed, eq.~(\ref{delsum0}) is 
obtained by substituting the transformation of eq.~(\ref{gengauge}) into
eq.~(\ref{ampform}) and requiring that the amplitude is invariant. One may 
further decompose the generalised gauge parameters $\Delta_i$ (assuming these
to be local) as~\cite{Bern:2010yg}
\begin{equation}
\Delta_i=\sum_{\alpha_i}\Delta_{i,\alpha_i}\,p_{\alpha_i}^2.
\label{deltaex}
\end{equation}
That is, the $\Delta_i$ associated with diagram $i$ can be expanded in terms
of the inverse propagators of this diagram. The $\Delta_i$ factors then move
contributions between diagrams by cancelling propagators, so that contributions
from one diagram can be absorbed in another. \\

Once a form of the amplitude satisfying BCJ duality has been found, a 
remarkable conjecture states that one can turn it into a gravity amplitude, in
a straightforward manner. By stripping off the colour factors in 
eq.~(\ref{ampform}) and replacing them with another set of numerators 
$\{\tilde{n}_i\}$, the {\it double copy conjecture} states 
that~\cite{Bern:2010ue,Bern:2010yg}\footnote{Note that in this paper we use
$\kappa=\sqrt{16\pi G_N}$ rather than $\kappa=\sqrt{32\pi G_N}$ (see 
section~\ref{sec:IRgrav}). This modifies the coupling factors in 
eq.~(\ref{ampform2}) relative to those in~\cite{Bern:2010ue,Bern:2010yg}.}
\begin{equation}
{\cal M}_m^{(L)}=i^{L+1}\left(\frac{\kappa}{\sqrt{2}}\right)^{m-2+2L}
\sum_{i\in\Gamma}\int\prod_{l=1}^L\frac{d^Dp_l}
{(2\pi)^D}\frac{1}{S_i}\frac{n_i\,\tilde{n}_i}{\prod_{\alpha_i}p_{\alpha_i}^2}
\label{ampform2}
\end{equation}
is an $m$-point, $L$-loop gravity amplitude. Note that the numerators $\{n_i\}$
and $\{\tilde{n}_i\}$ need not come from the same theory. In this paper, we
will be concerned with both sets of numerators coming from QCD, in which case
the gravity amplitude corresponds to general relativity (coupled to an
antisymmetric tensor and a dilaton). If the two gauge theories are ${\cal N}=N$
and ${\cal N}=M$ Super-Yang-Mills theory, then one obtains an amplitude in
${\cal N}=(N+M)$ supergravity. \\

This is all we need for what follows. We now turn to the study of infrared
singularities in QCD and gravity.

\subsection{Infrared singularities in non-Abelian gauge theory}
\label{sec:IRQCD}

Infrared singularities in quantum field theory have been studied over many 
decades, with a vast accompanying literature. Here we briefly summarise only 
those facts which are of direct relevance to what follows. For a more detailed
pedagogical review, see e.g.~\cite{Gardi:2009zv}. Whilst our statements can
be interpreted in the context of a general non-Abelian gauge theory, we 
explicitly refer to QCD throughout.\\

It is by now well-known that infrared singularities factorise, such that the 
general structure of an $m$-point scattering amplitude in QCD is, 
schematically, 
\begin{equation}
{\cal A}_m={\cal H}_m\cdot{\cal S}\cdot\prod_{i=1}^m\frac{J_i}{{\cal J}_i}.
\label{ampfac} 
\end{equation}
Here ${\cal H}_m$ is an infrared finite {\it hard interaction}, which is 
dressed by a universal soft function ${\cal S}$ that collects all infrared 
singularities. The {\it jet function} $J_i$ collects collinear singularities
associated with external leg $i$, and the {\it eikonal jet function} 
${\cal J}_i$ removes the double counting of divergences which are both soft
and collinear. The soft function is given by a vacuum expectation value of 
Wilson line operators along the space-time trajectories of the outgoing hard 
particles. Equivalently, one may calculate the soft part of a given hard
interaction by dressing all external lines with all possible soft gluon
emissions, according to the {\it eikonal Feynman rule} 
\begin{equation}
g_s{\bf T}_i\frac{p^\mu}{p\cdot k},
\label{eikrule}
\end{equation}
where $g_s$ is the strong coupling constant, and $p$ ($k$) the momentum of
the hard external line (soft gluon) respectively. Furthermore, ${\bf T}_i$
is a colour generator in the representation of external line $i$, where we have
used the notation of~\cite{Catani:1996jh,Catani:1996vz}. Where soft gluons
meet off the external lines, they couple according to the usual three and
four gluon vertices of QCD. \\

All singularities appearing in eq.~(\ref{ampfac}) can be shown to exponentiate,
so that instead of eq.~(\ref{ampfac}) we may write
\begin{equation}
{\cal A}_m={\cal H}_m\cdot Z,
\label{ampfac2}
\end{equation}
where 
\begin{equation}
Z=\exp\left[\sum_{n=1}^\infty c_n(\{p_i\},\epsilon,\mu)\alpha_S^n\right]
\label{Zdef}
\end{equation}
contains all soft and collinear singularities, and depends upon the momenta
$\{p_i\}$, as well as the dimensional regularisation parameter $\epsilon$
and scale $\mu$. The structure of the exponent (i.e. the form of the 
coefficients $\{c_n\}$ in eq.~(\ref{Zdef})) is known explicitly up to
two loop order for both massless and massive particles~\cite{Aybat:2006wq,
Aybat:2006mz,Kidonakis:2009ev,Mitov:2009sv,Becher:2009kw,Beneke:2009rj,
Czakon:2009zw,Ferroglia:2009ep,Ferroglia:2009ii,Chiu:2009mg,Mitov:2010xw}.
For massless particles, it has the remarkable property of involving both 
colour and kinematic correlations between at most pairs of particles, despite 
the fact that one would na\"{i}vely expect correlations between more than two 
particles to appear at two loop order and beyond. This property motivated the 
conjecture of the so-called {\it dipole formula} in QCD~\cite{Dixon:2008gr,
Becher:2009cu,Becher:2009qa,Gardi:2009qi,Gardi:2009zv}, which gives the 
all-order structure of eq.~(\ref{Zdef}) as
\begin{equation}
Z=\exp\left\{\int_0^{\mu^2}\frac{d\lambda^2}{\lambda^2}\left[\frac{1}{8}
\hat{\gamma}_K\left(\alpha_S(\lambda^2,\epsilon)\right)\sum_{(i,j)}\ln
\left(\frac{2p_i\cdot p_j\,e^{i\pi\lambda_{ij}}}{\lambda^2}\right){\bf T}_i
\cdot {\bf T}_j-\frac{1}{2}\sum_{i=1}^m\gamma_{J_i}\left(\alpha_S(\lambda^2,
\epsilon)\right)\right]\right\}.
\label{dipole}
\end{equation}
Here $\hat{\gamma}_K$ is the cusp anomalous dimension (itself a 
perturbative expansion in $\alpha_S$ with constant coefficients), and
$\gamma_{J_i}$ a further anomalous dimension associated with jet $i$, and
which collects hard collinear contributions. The double sum in the first term
is over all pairs of particles $(i,j)$, and following~\cite{Gardi:2009zv} we
have used the notation $\lambda_{ij}=1$ if $i$ and $j$ are both in the initial
or both in the final state (otherwise $\lambda_{ij}=0$). 
Equation~(\ref{dipole}) indeed contains correlations between dipoles only,
and is known to break down already at two loop order for massive external
legs. For the massless case, corrections may occur at three loop order and
beyond, either through explicit kinematic dependence of the relevant Feynman
integrals, or through higher order Casimir invariants appearing in the cusp
anomalous dimension. The form of possible corrections has been investigated 
in~\cite{Dixon:2009ur,Bret:2011xm,DelDuca:2011ae,Vernazza:2011aa,
Ahrens:2012qz}. Further progress may be made using recently developed 
techniques for classifying the structure of the exponent~\cite{Mitov:2010rp,
Gardi:2010rn,Gardi:2011wa,Gardi:2011yz}, or by considering alternative 
gauges~\cite{Chien:2011wz}.

\subsection{Infrared singularities in gravity}
\label{sec:IRgrav}

Complementary to the above mentioned studies in gauge theory, IR singularities
have also been investigated in gravity, commencing with the classic work 
of~\cite{Weinberg:1965nx}. Recently, there has been a revival of interest,
which has focused in particular on writing the structure of gravitational IR
divergences using the same language as is used in modern QCD studies. 
Reference~\cite{Naculich:2011ry} suggested the use of the following 
gravitational generalisation of eq.~(\ref{ampfac}):
\begin{equation}
{\cal M}_m={\cal H}_m\cdot{\cal S}^{\rm grav.},
\label{ampfacgrav}
\end{equation}
where ${\cal M}_m$ is an $m$-point gravity amplitude. Here ${\cal H}_m$ is 
again a hard interaction which is infrared finite, and ${\cal S}^{\rm grav.}$
is a universal gravitational soft function, which is given by a vacuum
expectation value of suitable Wilson line operators. There are no jet 
functions, due to the fact that collinear singularities cancel in gravity after
summing over all diagrams and using momentum 
conservation. The latter property was first established in the soft 
limit~\cite{Weinberg:1965nx}, but has recently been fully extended to encompass
hard collinear singularities~\cite{Akhoury:2011kq,Beneke:2012xa}. Furthermore,
ref.~\cite{White:2011yy} examined the form of eq.~(\ref{ampfacgrav}) in more 
detail using the path integral approach of~\cite{Laenen:2008gt}, also 
classifying what happens beyond the eikonal approximation. A similar structure
of next-to-eikonal corrections was found as in the case of 
gauge theory, as explored in~\cite{Laenen:2008gt,Laenen:2010uz}. Gravitational
Wilson lines in a soft-graviton context were further explored 
in~\cite{Miller:2012an} using the radial coordinate space picture 
of~\cite{Chien:2011wz}, and a continuous deformation was found between the 
cusp anomalous dimensions of QED and gravity at one loop. \\

As in the gauge theory case, the soft function in gravity exponentiates. 
However, a drastic simplification over gauge theory occurs in that the exponent
contains only one-loop diagrams, with no higher order corrections. This 
property is known as {\it one-loop exactness}, and has been firmly established
by the studies of~\cite{Weinberg:1965nx,Naculich:2011ry,Akhoury:2011kq}. It
implies that all infrared singularities (i.e. to all orders in perturbation
theory) ultimately stem from the exponentiation of the one-loop corrections,
in marked contrast to the gauge theory case of eq.~(\ref{Zdef}): even if the
dipole formula of eq.~(\ref{dipole}) happens to be true, there is still a
further perturbation expansion to be carried out in the exponent, whose soft 
part requires the cusp anomalous dimension to all orders. It is amusing to
note, as has already been mentioned in the introduction, that one-loop 
exactness in gravity implies that all infrared singularities are associated
with pairs of particles (the most that can be correlated with a single graviton
exchange). This is reminiscent of the QCD dipole formula in some sense (i.e.
that higher multipole correlations vanish), and allows us to speculate 
as to whether the dipole formula may have a gravitational origin. The results 
of this paper would appear to suggest that this is not the case, due to the 
disappearance of many singularities upon taking the double copy of a gauge 
theory. However, one-loop exactness has another important role: it tells us
that we know the all-order structure of IR singularities in gravity completely.
We can then ask whether the known IR divergence structures in QCD and gravity
are consistent with each other, if we apply the double copy procedure (via
BCJ duality). If this is so, this provides all-order evidence (at least in a 
particular limit) for the double copy conjecture. This will be the aim of the
rest of the paper.\\

In our subsequent calculations we will use the following conventions
(a number of different choices exist in the literature - see 
e.g.~\cite{Hamber:2007fk} for a convenient reference). The graviton field 
$h_{\mu\nu}$ is defined in terms of the metric tensor $g_{\mu\nu}$ via
\begin{equation}
g_{\mu\nu}=\eta_{\mu\nu}+\kappa h_{\mu\nu},
\label{hdef}
\end{equation}
where $\eta_{\mu\nu}$ is the Minkowski space metric, $\kappa=\sqrt{16\pi G_N}$
and $G_N$ is Newton's constant. Emission of soft gravitons with momentum $k$ 
from a hard external line of momentum $p$ is then given by the eikonal 
Feynman rule (see~\cite{Miller:2012an} for a recent derivation using the above 
conventions) 
\begin{equation}
\frac{\kappa}{2}\frac{p_\mu p_\nu}{p\cdot k},
\label{eikrulegrav}
\end{equation}
which may be compared with the gauge theory eikonal Feynman rule of 
eq.~(\ref{eikrule}). Finally, we will need the graviton propagator, for which
we use the result in the de Donder gauge:
\begin{equation}
D_{\mu\nu,\alpha\beta}(k)=\frac{-iP_{\mu\nu,\alpha\beta}}{k^2-i\epsilon},\quad
P_{\mu\nu,\alpha\beta}=\eta_{\mu\alpha}\,\eta_{\nu\beta}
+\eta_{\mu\beta}\,\eta_{\nu\alpha}-\frac{2}{D-2}\eta_{\mu\nu}\,
\eta_{\alpha\beta}. 
\label{prop}
\end{equation}
Having reviewed the necessary material for what follows, we begin our 
investigation of the double copy procedure in the soft limit in the following
section.

\section{One loop analysis}
\label{sec:1loop}
In the previous section, we reviewed various theoretical ideas concerning
BCJ duality, the double copy, and the structure of infrared singularities
in both QCD and gravity. This motivated the central question of our study: do
the known IR singularity structures in QCD and gravity match up with each other
upon taking the double copy of QCD? The aim of this paper is to argue that this
is indeed the case, and we will begin at one-loop order with the following 
strategy:
\begin{enumerate}
\item We classify all possible BCJ relations of the form of 
eq.~(\ref{jacobi2}), between sets of three diagrams containing cubic vertices,
in the case of pure Yang-Mills theory.
\item Next, we write down all possible diagrams which give infrared 
singularities at this order, and show that they can be matched up in sets of
three, each of which offers an explicit solution of a BCJ relation in the soft
limit. We will also see that additional BCJ relations cannot be explicitly 
satisfied (or otherwise) in the Feynman gauge in the soft limit, but that this 
will be irrelevant for the double copy. 
\item We then take the double copy of all infrared divergent diagrams, and verify that this reproduces the one-loop divergences in GR.
\end{enumerate}
The first step is to obtain the BCJ relations, which we do in the following
subsection.

\subsection{BCJ relations}
\label{sec:1loopBCJ}
\begin{figure}
\begin{center}
\scalebox{0.8}{\includegraphics{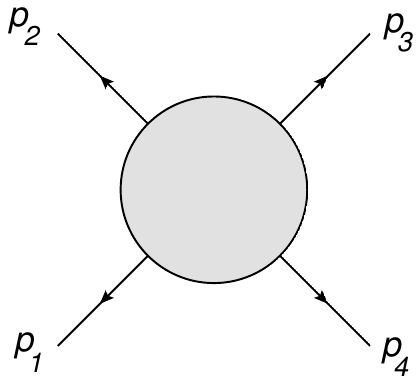}}
\caption{Four point scattering, where all momenta are taken to be outgoing.}
\label{fig:4point}
\end{center}
\end{figure}

\begin{figure}
\begin{center}
\scalebox{0.8}{\includegraphics{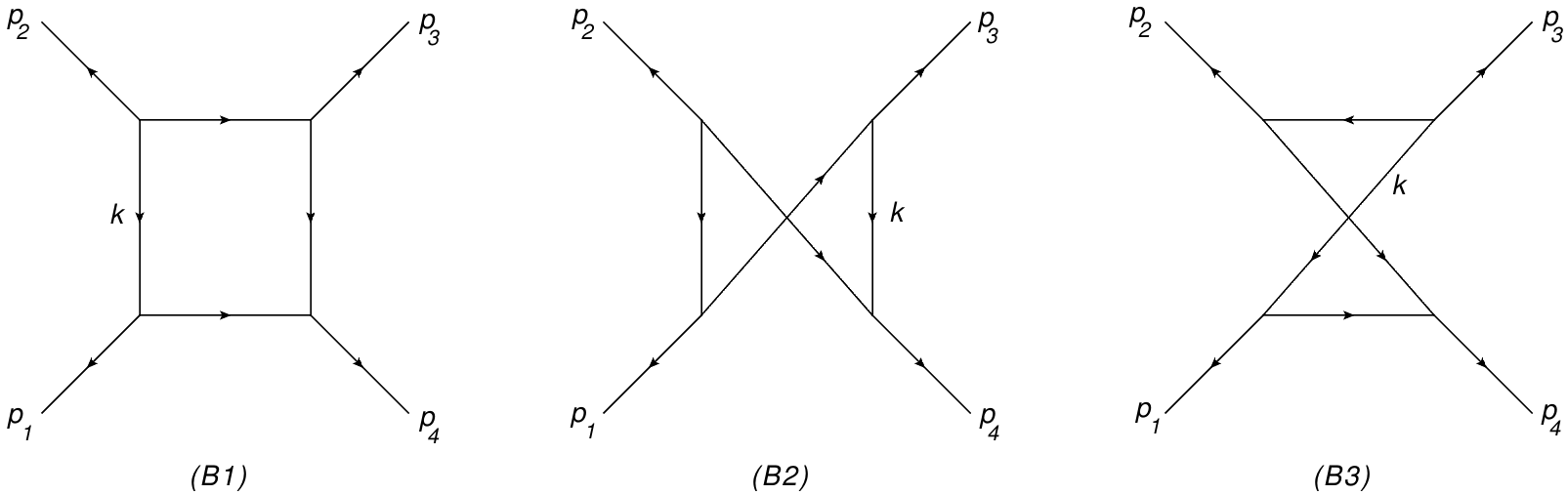}}
\caption{Set of all box graphs at one-loop order, where we depict the basis momenta entering the numerator for each graph.}
\label{fig:boxes}
\end{center}
\end{figure}

\begin{figure}
\begin{center}
\scalebox{0.8}{\includegraphics{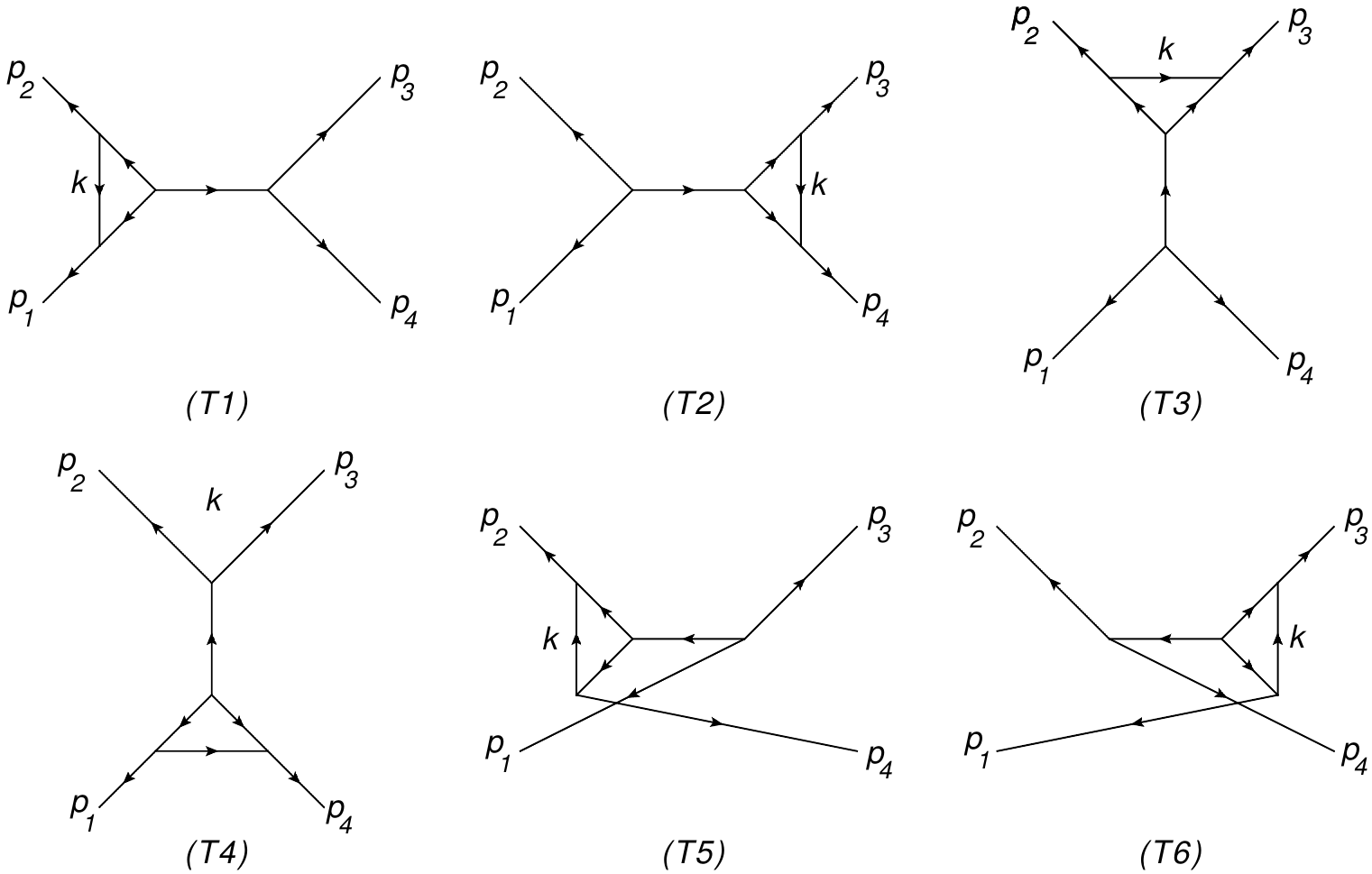}}
\caption{Set of all triangle graphs at one-loop order, where we depict the basis momenta entering the numerator for each graph.}
\label{fig:triangles}
\end{center}
\end{figure}

\begin{figure}
\begin{center}
\scalebox{0.8}{\includegraphics{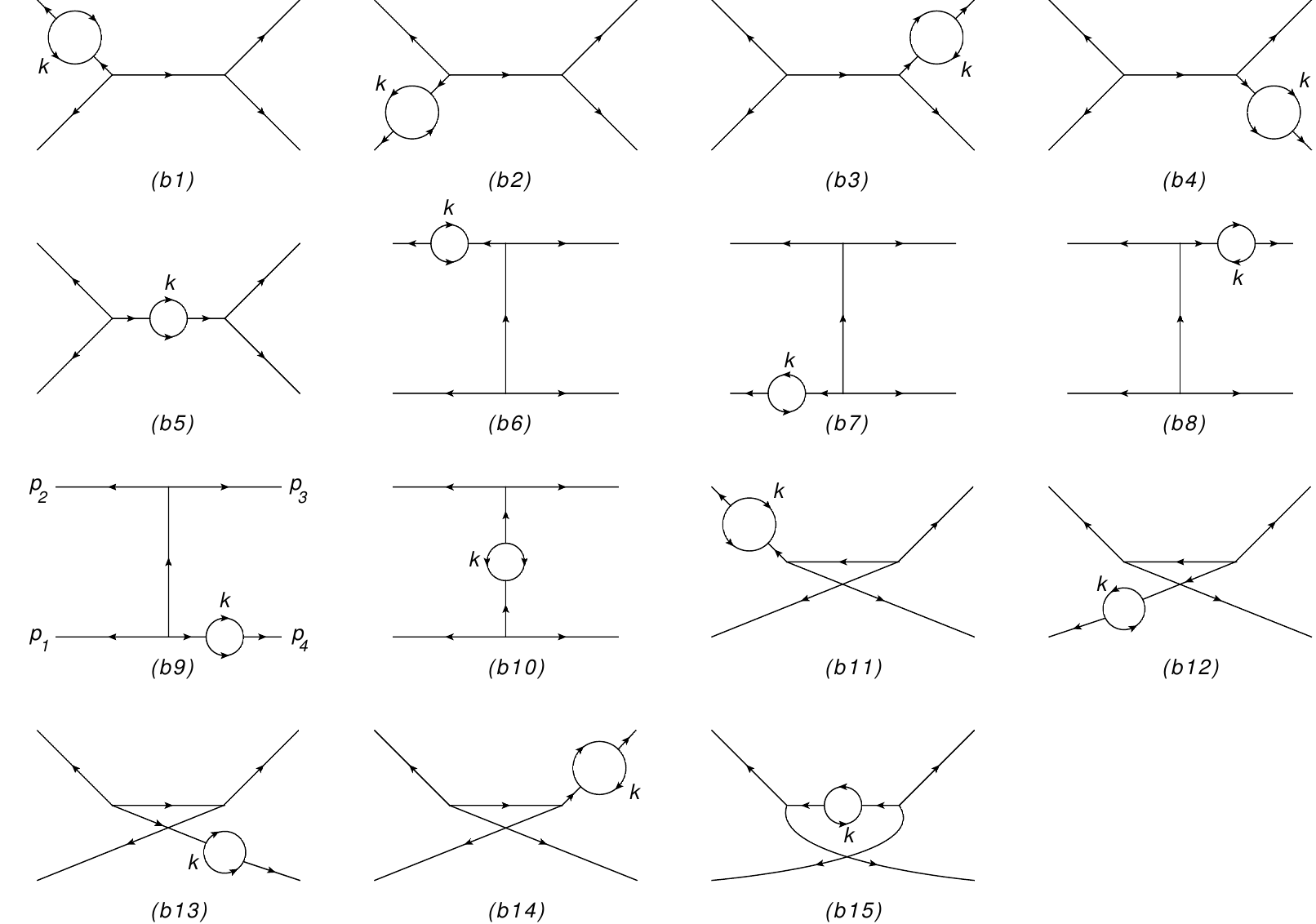}}
\caption{Set of all bubble graphs at one-loop order, where we show the choice of internal basis momentum for each graph. All external momenta are labelled as in graph (b9).}
\label{fig:bubbles}
\end{center}
\end{figure}

To start with, we must write down all possible one-loop graphs, and for a 
concrete example we consider the case of 4-point scattering, 
labelled as in figure~\ref{fig:4point}. We take all momenta outgoing, and 
define Mandelstam invariants according to
\begin{equation}
s=(p_1+p_2)^2,\quad t=(p_1+p_4)^2,\quad u=(p_1+p_3)^2,
\label{mandies}
\end{equation}
for ease of comparison with e.g.~\cite{Bern:2008qj}. For each graph, we specify
the orientation of momentum on each line, together with a set of basis momenta
for the corresponding numerator function (three external, and one internal). 
All other momenta in each graph can then be fixed from these momenta. The box,
triangle and bubble graphs are shown in figures~\ref{fig:boxes}, 
\ref{fig:triangles} and \ref{fig:bubbles} respectively.\\

As explained in section~\ref{sec:review}, the diagrams in 
figures~\ref{fig:boxes}-\ref{fig:bubbles} form sets of three,
related by subjecting internal lines to a BCJ transformation (replacement
of $t$-channel-like exchange by $s$- and $u-$channel exchanges). The colour
factors of such a set satisfy the Jacobi identity, and BCJ duality, if 
satisfied, then requires a corresponding set of functional equations to hold 
for the kinematic numerators associated with each graph. The BCJ relations in
the present case can be divided into four classes. Firstly, there are relations
which relate two box topologies to a triangle. These are (using the labels and
basis momenta from figures~\ref{fig:boxes} and~\ref{fig:triangles})
\begin{align}
n_{T1}(p_1,p_2,p_3,k)&=n_{B1}(p_1,p_2,p_3,k)-n_{B2}(p_1,p_2,p_3,-p_1-p_3+k);\notag\\
n_{T2}(p_1,p_2,p_3,k)&=n_{B1}(p_1,p_2,p_3,-p_2-p_3-k)-n_{B2}(p_1,p_2,p_3,k);\notag\\
n_{T3}(p_1,p_2,p_3,k)&=n_{B1}(p_1,p_2,p_3,-k-p_2)-n_{B3}(p_1,p_2,p_3,-p_3+k);\notag\\
n_{T4}(p_1,p_2,p_3,k)&=n_{B1}(p_1,p_2,p_3,k+p_1)-n_{B3}(p_1,p_2,p_3,k+p_1);\notag\\
n_{T5}(p_1,p_2,p_3,k)&=n_{B3}(p_1,p_2,p_3,k-p_2-p_3)-n_{B2}(p_1,p_2,p_3,k-p_1-p_2-p_3);\notag\\
n_{T6}(p_1,p_2,p_3,k)&=n_{B3}(p_1,p_2,p_3,-k)-n_{B2}(p_1,p_2,p_3,k-p_3).
\label{BCJBBT}
\end{align}
We have chosen to write these in terms of the triangles evaluated with loop
momentum $k$ as the fourth argument, for reasons that will become clear.
Next, there are BCJ relations that relate each triangle topology to the same
topology (evaluated with a different loop momentum argument) and a 
bubble. There are two subclasses - firstly those relations which involve an
external self-energy:
\begin{align}
n_{T1}(p_1,p_2,p_3,k)&=-n_{T1}(p_1,p_2,p_3,-p_2-k)+n_{b1}(p_1,p_2,p_3,k)\notag\\
n_{T1}(p_1,p_2,p_3,k)&=-n_{T1}(p_1,p_2,p_3,p_1-k)+n_{b2}(p_1,p_2,p_3,k)\notag\\
n_{T2}(p_1,p_2,p_3,k)&=-n_{T2}(p_1,p_2,p_3,-p_3-k)+n_{b3}(p_1,p_2,p_3,k)\notag\\
n_{T2}(p_1,p_2,p_3,k)&=-n_{T2}(p_1,p_2,p_3,-p_1-p_2-p_3-k)+n_{b4}(p_1,p_2,p_3,k)\notag\\
n_{T3}(p_1,p_2,p_3,k)&=-n_{T3}(p_1,p_2,p_3,-p_2-k)+n_{b6}(p_1,p_2,p_3,k)\notag\\
n_{T3}(p_1,p_2,p_3,k)&=-n_{T3}(p_1,p_2,p_3,p_3-k)+n_{b8}(p_1,p_2,p_3,k-p_3)\notag\\
n_{T4}(p_1,p_2,p_3,k)&=-n_{T4}(p_1,p_2,p_3,-p_1-k)+n_{b7}(p_1,p_2,p_3,k+p_1)\notag\\
n_{T4}(p_1,p_2,p_3,k)&=-n_{T4}(p_1,p_2,p_3,-p_1-p_2-p_3-k)+n_{b9}(p_1,p_2,p_3,-p_1-p_2-p_3-k)\notag\\
n_{T5}(p_1,p_2,p_3,k)&=-n_{T5}(p_1,p_2,p_3,p_2-k)+n_{b11}(p_1,p_2,p_3,k-p_2)\notag\\
n_{T5}(p_1,p_2,p_3,k)&=-n_{T5}(p_1,p_2,p_3,p_1+p_2+p_3-k)+n_{b13}(p_1,p_2,p_3,-k);\notag\\
n_{T6}(p_1,p_2,p_3,k)&=-n_{T6}(p_1,p_2,p_3,-p_1-k)+n_{b12}(p_1,p_2,p_3,p_1+k)\notag\\
n_{T6}(p_1,p_2,p_3,k)&=-n_{T6}(p_1,p_2,p_3,p_3-k)+n_{b14}(p_1,p_2,p_3,-k).
\label{BCJTTb}
\end{align}
Secondly, there are those which involve an internal self-energy:
\begin{align}
n_{T1}(p_1,p_2,p_3,k)&=-n_{T1}(p_1,p_2,p_3,p_1-p_2-k)+n_{b5}(p_1,p_2,p_3,-p_2-k)\notag\\
n_{T2}(p_1,p_2,p_3,k)&=-n_{T2}(p_1,p_2,p_3,-p_1-p_2-2p_3-k)+n_{b5}(p_1,p_2,p_3,p_3+k)\notag\\
n_{T3}(p_1,p_2,p_3,k)&=-n_{T3}(p_1,p_2,p_3,p_3-p_2-k)+n_{b10}(p_1,p_2,p_3,-p_2-k)\notag\\
n_{T4}(p_1,p_2,p_3,k)&=-n_{T4}(p_1,p_2,p_3,-2p_1-p_2-p_3-k)+n_{b10}(p_1,p_2,p_3,k+p_1)\notag\\
n_{T5}(p_1,p_2,p_3,k)&=-n_{T5}(p_1,p_2,p_3,p_1+2p_2+p_3-k)+n_{b15}(p_1,p_2,p_3,-k+p_1+p_2+p_3)\notag\\
n_{T6}(p_1,p_2,p_3,k)&=-n_{T6}(p_1,p_2,p_3,-p_1+p_3-k)+n_{b15}(p_1,p_2,p_3,p_1+k). 
\label{BCJTTb2}
\end{align}
Next, there are relations which relate the numerators of three bubble graphs:
\begin{align}
n_{b1}(p_1,p_2,p_3,k)&=n_{b6}(p_1,p_2,p_3,-p_2-k)+n_{b11}(p_1,p_2,p_3,-p_2-k)\notag\\
n_{b2}(p_1,p_2,p_3,k)&=n_{b7}(p_1,p_2,p_3,k)+n_{b12}(p_1,p_2,p_3,k)\notag\\
n_{b3}(p_1,p_2,p_3,k)&=n_{b8}(p_1,p_2,p_3,k)+n_{b14}(p_1,p_2,p_3,k)\notag\\
n_{b4}(p_1,p_2,p_3,k)&=n_{b9}(p_1,p_2,p_3,k)+n_{b13}(p_1,p_2,p_3,-p_1-p_2-p_3-k).
\label{BCJbbb}
\end{align}
Each of these has the form of a complete set of $s-$, $t-$ or $u$-channel 
tree-level graphs, dressed by a common external self-energy.
Finally, there are relations which relate two bubble graphs with a tadpole. 
Here we assume that tadpole graphs can be set to zero, in which case the 
remaining relations impose constraints on the forms of the bubble numerators
as follows:
\begin{equation}
n_{bi}(p_1,p_2,p_3,k)=n_{bi}(p_1,p_2,p_3,-k)\quad\forall\quad1\leq i\leq15.
\label{BCJbb}
\end{equation}
A similar set of BCJ relations has been presented recently 
in~\cite{Carrasco:2012ca}, in the context of ${\cal N}=1$ and ${\cal N}=2$ SYM
theory. However, in that case it was found that the bubble numerators could be
set to zero.\\

Having presented the one-loop relations in this section, we proceed to consider
the soft limit. Our aim is to show that the BCJ relations are satisfied in 
this limit, up to corrections which are subleading in soft momentum, and which
are irrelevant for the double copy to gravity. This is the subject of the 
following section. 

\subsection{Soft limit}
\label{sec:1loopsoft}

In this section, we consider all possible infrared singular diagrams at
one loop order, and show that their numerators are consistent with the BCJ
relations obtained in the previous section. A full solution of all relations
requires the inclusion of higher order terms in the soft loop momentum. 
However, these corrections will be seen to be irrelevant in taking the double
copy to gravity, if one is focusing solely on infrared divergences. \\

\begin{figure}
\begin{center}
\scalebox{0.8}{\includegraphics{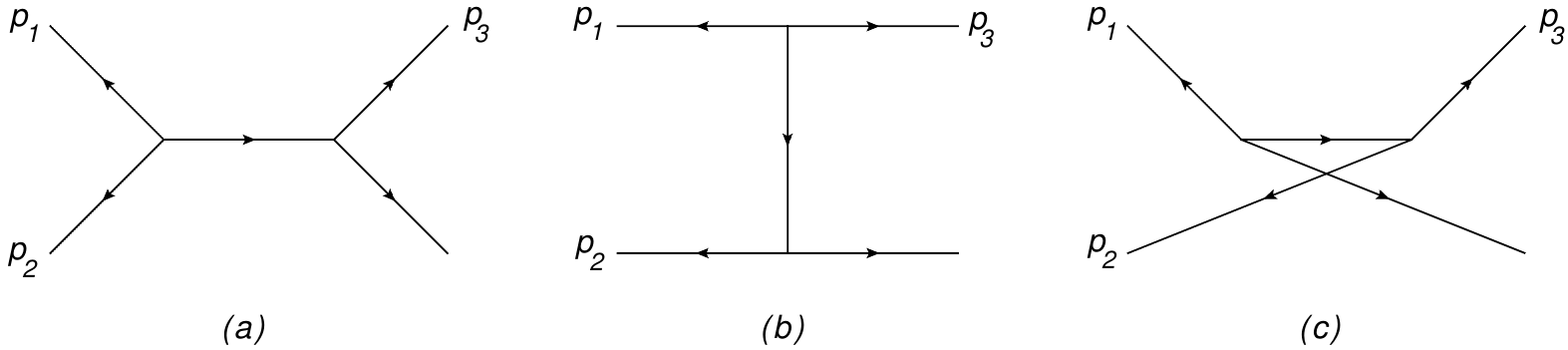}}
\caption{Labelling of basis momenta for the tree-level numerators $n_x(p_1,p_2,p_3)$.}
\label{fig:treelevel}
\end{center}
\end{figure}

IR singularities at one loop are generated by dressing tree level diagrams with
soft gluon emissions. Considering once again the concrete example of 4-point 
scattering, we may write the full tree level amplitude as\footnote{Note that 
we here adopt the phase conventions of eq.~(\ref{ampform}).
Associating a factor $i$ and $-i$ with each propagator and vertex respectively,
one may absorb a further factor of $1/i$ in the numerators $n_x$ to obtain an
overall power of $i^L$ at $L$-loop order. Performing the double copy involves
an extra numerator, and thus an additional explicit factor of $i$ in the 
gravity amplitude.}
\begin{equation}
{\cal A}^{(0)}=\sum_{x\in\{s,t,u\}}{\cal A}_x,\quad {\cal A}_x=\frac{c_xn_x(p_1,p_2,p_3)}{x},
\label{A0def}
\end{equation}
where $\{n_x\}$ is a suitable set of numerators (with colour factors $\{c_i\}$
) obeying the tree-level BCJ relations
\begin{equation}
n_s(p_1,p_2,p_3)-n_t(p_1,p_2,p_3)-n_u(p_1,p_2,p_3)=0,\quad c_s-c_t-c_u=0.
\label{BCJtree}
\end{equation}
Note that each tree-level numerator requires three basis momenta, and we label
these according to figure~\ref{fig:treelevel} (n.b. the choice of external 
momenta is similar to that used in figures~\ref{fig:boxes}-\ref{fig:bubbles}).
Note that we have absorbed coupling factors into the tree level numerators, so
that all powers of the coupling in subsequent equations correspond to higher 
order corrections. Also, implicit in eq.~(\ref{A0def}) is the fact that the 
four-gluon vertex graph has been rewritten in terms of cubic graphs. 
The effect of dressing a tree-level amplitude ${\cal A}_x$ with a soft gluon 
exchanged between legs $i$ and $j$ is given by
\begin{displaymath}
g_s^2{\bf T}_i\cdot{\bf T}_j\,{\cal I}_{ij}\,{\cal A}_x,
\end{displaymath}
where the eikonal integral factor
\begin{equation}
{\cal I}_{ij}=i\int\frac{d^D k}{(2\pi)^D}\frac{p_i\cdot p_j}
{k^2\,p_i\cdot k\,p_j\cdot k},
\label{Idef}
\end{equation}
as results from connecting two eikonal Feynman rules (eq.~(\ref{eikrule})) 
with a gluon propagator (note we use the Feynman gauge). The full infrared
singular part of the one-loop amplitude can then be 
written~\footnote{Strictly speaking the eikonal integral of eq.~(\ref{Idef}) is
zero, being a scaleless integral in dimensional regularisation. However, this
is due to the cancellation of the IR divergence with a spurious UV pole which 
results from the replacement of quadratic propagators by linear ones, such that
one may recover the IR divergence using an additional regularisation 
procedure.}
\begin{equation}
{\cal A}^{(1)}_{\rm IR}=ig_s^2\sum_x\sum_{i<j}{\bf T}_i\cdot{\bf T}_{j}\,
{\cal I}_{ij}\frac{{\cal A}_x}{x}.
\label{A1IRdef}
\end{equation}
Here the first sum is over the three tree-level cubic topologies, and the 
second sum is over all pairs of external legs, where each pair is counted only
once. There are no contributions from soft emissions which begin and end
on the same external line, due to the fact that all outgoing hard particles
are massless ($p_i^2=0$). Furthermore, there are no contributions from 
internal self-energy diagrams: such graphs contain additional hard propagators
that remove the infrared singularity. \\

\begin{figure}
\begin{center}
\scalebox{0.8}{\includegraphics{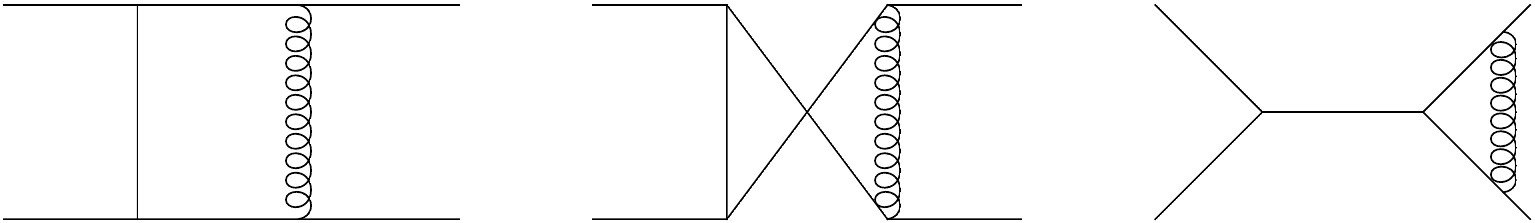}}
\caption{Diagrams involving a soft gluon exchange between legs 3 and 4, where
solid (curly) lines denote hard (soft) gluons respectively, and we have suppressed the orientations on all lines for brevity.}
\label{fig:12soft}
\end{center}
\end{figure}

We can now interpret eq.~(\ref{A1IRdef}) from the viewpoint of BCJ duality. 
First, let us examine the relations of eq.~(\ref{BCJBBT}), relating pairs of
box graphs with a triangle. These relations may be satisfied by grouping the
terms appearing in eq.~(\ref{A1IRdef}) into sets of three. Interchanging
the orders of summation in eq.~(\ref{A1IRdef}), one may pick out a particular
pair $(i,j)$, to give
\begin{equation}
\sum_x g_s^2{\bf T}_i\cdot{\bf T}_{j}\int\frac{d^Dk}{(2\pi)^D}
\frac{p_i\cdot p_j}{k^2\,p_i\cdot k\,p_j\cdot k}\frac{c_x n_x(p_1,p_2,p_3)}{x},
\label{termij}
\end{equation}
where we have substituted in the explicit forms for ${\cal I}_{ij}$ and 
${\cal A}_x$ from eqs.~(\ref{A0def}) and~(\ref{Idef}). This is three 
separate terms, each of which can be interpreted as a soft limit of one of
the graphs appearing in figures~\ref{fig:boxes}-\ref{fig:triangles}. 
Taking the pair (3,4) as an example, we may interpret eq.~(\ref{termij}) as 
shown in figure~\ref{fig:12soft}. 
From eq.~(\ref{termij}), we may associate with each term a collected numerator
and colour factor~\footnote{Note that the colour generators ${\bf T}_i$ are 
understood with indices ordered in the direction of momentum flow on each 
external line.}. 
\begin{equation}
\hat{n}_{x,ij}=(p_i\cdot p_j)n_x(p_1,p_2,p_3),\quad c_{x,ij}={\bf T}_i\cdot{\bf T}_j\,c_x,
\label{nxijdef}
\end{equation} 
To make this mapping more precise, we may identify the soft limits of the 
numerator functions appearing in eq.~(\ref{BCJBBT}) as follows. For the 
triangles, one has
\begin{align}
n_{T1}(p_1,p_2,p_3,k)&=\hat{n}_{s,12}+{\cal O}(k);\notag\\
n_{T2}(p_1,p_2,p_3,k)&=\hat{n}_{s,34}+{\cal O}(k);\notag\\
n_{T3}(p_1,p_2,p_3,k)&=\hat{n}_{t,23}+{\cal O}(k);\notag\\
n_{T4}(p_1,p_2,p_3,k)&=\hat{n}_{t,14}+{\cal O}(k);\notag\\
n_{T5}(p_1,p_2,p_3,k)&=\hat{n}_{u,24}+{\cal O}(k);\notag\\
n_{T6}(p_1,p_2,p_3,k)&=\hat{n}_{u,13}+{\cal O}(k).
\label{trilims}
\end{align}
For the boxes, there is more than one soft limit for each graph, corresponding
to the fact that there is an infrared singular region when any of the internal
lines has vanishing momentum. We may express these limits as~\footnote{Note 
that care is needed regarding minus signs in eq.~(\ref{boxlims}). The sign of
each soft limit is fixed by the collected colour factor in eq.~(\ref{nxijdef}),
and its comparison with figures~\ref{fig:boxes} and~\ref{fig:treelevel}.}
\begin{align}
&n_{B1}(p_1,p_2,p_3,k)=\hat{n}_{t,12}+{\cal O}(k);\notag\\
&n_{B1}(p_1,p_2,p_3,-p_2-p_3-k)=\hat{n}_{t,34}+{\cal O}(k);\notag\\
&n_{B1}(p_1,p_2,p_3,-k-p_2)=\hat{n}_{s,23}+{\cal O}(k);\notag\\
&n_{B1}(p_1,p_2,p_3,k+p_1)=\hat{n}_{s,14}+{\cal O}(k);\notag\\
&n_{B2}(p_1,p_2,p_3,k)=-\hat{n}_{u,34}+{\cal O}(k);\notag\\
&n_{B2}(p_1,p_2,p_3,-p_1-p_3+k)=-\hat{n}_{u,12}+{\cal O}(k);\notag\\
&n_{B2}(p_1,p_2,p_3,k-p_1-p_2-p_3)=-\hat{n}_{s,24}+{\cal O}(k);\notag\\
&n_{B2}(p_1,p_2,p_3,k-p_3)=-\hat{n}_{s,13}+{\cal O}(k);\notag\\
&n_{B3}(p_1,p_2,p_3,k)=-\hat{n}_{t,13}+{\cal O}(k);\notag\\
&n_{B3}(p_1,p_2,p_3,-p_3+k)=\hat{n}_{u,23}+{\cal O}(k);\notag\\
&n_{B3}(p_1,p_2,p_3,k+p_1)=\hat{n}_{u,14}+{\cal O}(k);\notag\\
&n_{B3}(p_1,p_2,p_3,k-p_2-p_3)=-\hat{n}_{t,24}+{\cal O}(k).\notag\\
\label{boxlims}
\end{align}
One may then verify the relations
\begin{equation}
\hat{n}_{s,ij}-\hat{n}_{u,ij}-\hat{n}_{t,ij}=0,\quad
c_{s,ij}-c_{u,ij}-c_{t,ij}=0,
\label{nxijBCJ}
\end{equation}
which follow from the fact that the tree level colour factors and numerators
satisfy eq.~(\ref{BCJtree}), and that these are multiplied by a common
factor in eq.~(\ref{nxijdef}). Equation~(\ref{nxijBCJ}) contains six different 
relations (the number of ways of choosing two external legs out of four), 
which correspond precisely to soft limits of the BCJ relations in 
eq.~(\ref{BCJBBT}). Take, for example, the case shown in 
figure~(\ref{fig:12soft}). These diagrams consist of the $k\rightarrow0$ 
limits of $n_{B1}(p_1,p_2,p_3,-p_2-p_3-k)$, $n_{B2}(p_1,p_2,p_3,k)$ and 
$n_{T3}(p_1,p_2,p_3,k)$, and the relevant BCJ relation from eq.~(\ref{nxijBCJ})
thus corresponds to the second relation in eq.~(\ref{BCJBBT}). \\

We now consider the remaining BCJ relations. It has already been remarked upon
above that bubble graphs do not contribute to the infrared singular part of the
one-loop scattering amplitude: the eikonal Feynman rules ensure that external
self-energies vanish, and internal bubbles vanish due to the fact that a 
gluon cuts a hard internal propagator. It is equally true that bubble 
topologies which result from rewriting four-gluon vertex graphs in terms of
bubble topologies ({\it snail graphs} in the language of 
e.g.~\cite{Bern:2012uf}) are not infrared singular. Thus, in the soft limit,
we may set
\begin{equation}
\hat{n}_{bi}={\cal O}(k) \quad\forall\quad i,
\label{hatnbi}
\end{equation}
where the hat above the numerator implies that we are evaluating this only in 
the soft limit. Such numerators are defined only up to arbitrary terms
containing at least one power of a soft gluon momentum $k$, which in principle
must come from a full solution of the BCJ relations linking bubbles with other
topologies. However, such terms are irrelevant for the IR singularities of the
amplitude, and certainly once the double copy to gravity is taken. In the copy,
the numerator of each topology is squared, whilst the denominator remains 
unchanged. Thus, bubble graphs will also not contribute IR singularities on 
the gravity side. \\

With the bubble numerators set to zero, one does not have to worry about the
relations of eqs.~(\ref{BCJbbb}) and eq.~(\ref{BCJbb}). However, one must still
address the relations of eq.~(\ref{BCJTTb}), which link each triangle numerator
to the same numerator function (evaluated with different loop momenta), and a 
bubble. For these, the soft limit alone provides insufficient information as 
to whether such relations are satisfied. Clearly the full $k$ dependence 
(represented by the ${\cal O}(k)$ terms in eqs.~(\ref{trilims}) 
and~(\ref{hatnbi})) is necessary in order to verify these relations, and thus 
we are not able to solve these relations explicitly. However, this is again 
irrelevant for taking the double copy. The triangle numerators above are 
${\cal O}(k^0)$, whereas the bubble numerators are ${\cal O}(k)$. In modifying 
them by a generalised gauge transformation, they may potentially change by 
superpositions of denominator factors as given by eq.~(\ref{deltaex}). Given 
that all relevant denominators are at least ${\cal O}(k)$, this means that 
denominators which fully satisfy the BCJ relations have the same limits as 
given above. Corrections $\sim{\cal O}(k)$ will not give rise to additional
infrared singularities, and this remains true after taking the double copy 
given that the denominators do not change.\\

We have now seen that of the BCJ relations derived in the previous section
(for the case of full QCD), it is possible to identify the soft limits of 
the numerator functions, such that these satisfy the relations linking boxes
with triangles. For the remaining relations, the bubbles can be set to zero
given that they do not give rise to IR singularities. Furthermore, the 
relations linking triangles with bubbles imply corrections to the leading
soft behaviour that are irrelevant as far as infrared singularities are 
concerned.\\

Armed with the above knowledge, we are now permitted to take the double copy of
eq.~(\ref{A1IRdef}) as prescribed in~\cite{Bern:2010ue}, to 
give~\footnote{Here we have again adopted the phase conventions of 
eq.~(\ref{ampform}) - see the footnote on p. 12.}
\begin{align}
{\cal M}^{(1)}_{\rm IR}&=-\sum_x\sum_{i<j}\left(\frac{\kappa}{\sqrt{2}}
\right)^2\int\frac{d^Dk}{(2\pi)^D}\frac{(p_i\cdot p_j)^2}{k^2\,p_i\cdot k\,
p_j\cdot k}\frac{n_x n_x}{x}\notag\\
&=-\sum_x\sum_{i<j}\left(\frac{\kappa}{2}\right)^2
\int\frac{d^Dk}{(2\pi)^D}\frac{2(p_i\cdot p_j)^2}{k^2\,p_i\cdot k\,p_j\cdot k}
\frac{n_x n_x}{x},
\label{M1def}
\end{align}
where we write $n_x\equiv n_x(p_1,p_2,p_3)$.
If the double copy procedure is valid, this result should give the infrared 
singular parts of the 4-point, 1 loop amplitude in GR. That this is indeed the
case can be seen as follows. Firstly, the factor
\begin{displaymath}
\frac{in_xn_x}{x}
\end{displaymath}
is, by the tree-level double copy procedure, a gravitational tree level graph
for the $s$, $t$ or $u$ topology~\footnote{Recall that we absorbed gauge theory
coupling factors into the tree-level numerators. Implicit in eq.~(\ref{M1def})
is that these have been replaced appropriately in the double copy to gravity.}.
This is then dressed by an eikonal integral, which corresponds to the
appropriate gravitational generalisation of eq.~(\ref{Idef}). To check this,
one may contract eikonal Feynman rules on legs $i$ and $j$ with the propagator
of eq.~(\ref{prop}) to get
\begin{align}
{\cal I}_{ij}^{\rm grav.}&=\left(\frac{\kappa}{2}\right)^2\int\frac{d^D k}
{(2\pi)^D}\frac{p_i^\mu\,p_i^\nu}{p_i\cdot k}\left(-\frac{p_j^\mu\,p_j^\nu}
{p_j\cdot k}\right)\left(-\frac{i}{k^2}\right)\left(\eta_{\mu\alpha}
\eta_{\nu\beta}+\eta_{\mu\beta}\eta_{\nu\alpha}-\frac{2}{D-2}\eta_{\mu\nu}
\eta_{\alpha\beta}\right)\notag\\
&=i\left(\frac{\kappa}{2}\right)^2\int\frac{d^Dk}{(2\pi)^D}
\frac{2(p_i\cdot p_j)^2}{k^2\,p_i\cdot k\,p_j\cdot k},
\label{Iijcalc}
\end{align}
where the minus sign in the second eikonal Feynman rule results from reversing
the sign of the soft gluon momentum. Dressing the tree-level gravitational
interaction with equation~(\ref{Iijcalc}) is in exact agreement with 
eq.~(\ref{M1def}). \\

Note that, as first observed in~\cite{Weinberg:1965nx}, soft collinear
singularities cancel after summing over all diagrams. Taking those which
are associated with line $i$ as an example, these are generated by the sum
of eikonal integrals
\begin{align}
\sum_{j\neq i}\int\frac{d^Dk}{(2\pi)^D}\lim_{k\rightarrow p_i}\left[
\frac{4(p_i\cdot p_j)^2}{k^2\,p_i\cdot k\,p_j\cdot k}\right]&=
\sum_{j\neq i}\int\frac{d^Dk}{(2\pi)^D}\lim_{k\rightarrow p_i}\left[
\frac{4(p_i\cdot p_j)}{k^2\,p_i\cdot k}\right].
\label{col1}
\end{align}
Applying momentum conservation
\begin{equation}
p_i=-\sum_{j\neq i}p_j
\label{momcon}
\end{equation}
then gives zero on the right-hand side of eq.~(\ref{col1}), up to non-singular 
terms. Note that the QCD equivalent of eq.~(\ref{col1}) (also including colour 
factors) is
\begin{align}
\sum_{j\neq i}{\bf T}_i\cdot{\bf T}_j\int\frac{d^Dk}{(2\pi)^D}
\lim_{k\rightarrow p_i}\left[\frac{(p_i\cdot p_j)
}{k^2\,p_i\cdot k\,p_j\cdot k}\right]&=
\sum_{j\neq i}{\bf T}_i\cdot{\bf T}_j\int\frac{d^Dk}{(2\pi)^D}
\lim_{k\rightarrow p_i}\left[\frac{1}{k^2\,p_i\cdot k}\right].
\label{col2}
\end{align}
Colour conservation
\begin{equation}
{\bf T}_i=-\sum_{j\neq i}{\bf T}_j
\label{colcon}
\end{equation}
then gives
\begin{equation}
\sum_{j\neq i}{\bf T}_i\cdot{\bf T}_j\int\frac{d^Dk}{(2\pi)^D}
\lim_{k\rightarrow p_i}\left[\frac{(p_i\cdot p_j)
}{k^2\,p_i\cdot k\,p_j\cdot k}\right]=
C_i\int\frac{d^Dk}{(2\pi)^D}
\lim_{k\rightarrow p_i}\left[\frac{1}{k^2\,p_i\cdot k}\right],
\label{col3}
\end{equation}
where $C_i$ is the quadratic Casimir associated with external line $i$
($C_i=C_A=N_c$ for pure gluodynamics). This remains singular, and a simple
physical way to interpret this is that collinear singularities can only
depend on the quantum numbers of a single particle, which include the 
squared charge of the appropriate external line. In
QCD, this is the quadratic Casimir operator in the relevant representation,
whereas in gravity this is the squared 4-momentum, which vanishes if collinear
singularities are to be present (i.e. if the leg is massless). The double copy
procedure has here provided an interesting mechanism for the cancellation of
soft-collinear singularities on the gravity side: the colour dependence in QCD
gets replaced by additional momentum factors, which generate the necessary
squared 4-momentum. Note that this means that there are singularities on the
gauge theory side that vanish upon performing the double copy. We will see
this happening more generally at two-loop order and beyond.\\

We have now verified that the soft limit of GR is precisely reproduced upon
taking the double copy of the soft limit of QCD at one-loop order. Although
we here focused on the particular case of 4-point scattering, the argument
is easily generalised to any number of external legs. Having examined the
one-loop case, we proceed to two loops in the following section.

\section{Two loop analysis}
\label{sec:2loop}

In the previous section, we showed that the known infrared singularities of
GR are reproduced by double copying those of QCD at one-loop order. Application
of the double copy relied upon the fact that the BCJ relations could be 
satisfied in the soft limit. The solution of relations linking boxes with 
triangles was rather straightforward at this order, and relied ultimately on 
the fact that corresponding relations were satisfied (after removal of a common
soft gluon) at tree level. Put another way, the soft BCJ relations of
eq.~(\ref{nxijBCJ}) all had the form of a common eikonal factor 
$(p_i\cdot p_j)$ multiplying a tree-level relation. Further relations involving
at least one bubble graph could not be satisfied explicitly by the Feynman 
gauge results in the soft limit. However, this was irrelevant given that bubble
graphs could be consistently set to zero as far as IR singularities were 
concerned. Furthermore, corrections to triangle numerators were subleading in
the soft limit, and thus irrelevant for IR singularities, on both the gauge
theory and gravity sides of the double copy.

This simple structure will not quite be the case at two loop order. We will 
again see that numerators for individual graphs, computed in the Feynman gauge,
do not automatically satisfy BCJ relations in the soft limit. This will 
now affect IR singularities in the gauge theory, but will remain irrelevant for
the double copy, and hence for reproducing the known infrared singularities of 
GR. \\

Our strategy will be the same as at one-loop order, and begins by writing down
the BCJ relations in full QCD. Then one draws all possible soft topologies, 
and demonstrates that the resulting numerators satisfy BCJ relations in the
soft limit. Our experience at one loop tells us that we do not have to worry 
about most of the bubble diagrams. The numerators for these (whilst non-zero
in full QCD) can be set to zero in the soft limit as before: most external
and internal bubbles are IR-suppressed. The only exception to this is the
presence of self-energies associated with soft gluons, as we will see in what 
follows. There is a large number of diagrams at two-loop order and we do not
collect them all here. Rather, we consider directly relevant soft topologies, 
and show that these can indeed be matched up with BCJ relations.\\

\begin{figure}
\begin{center}
\scalebox{1.0}{\includegraphics{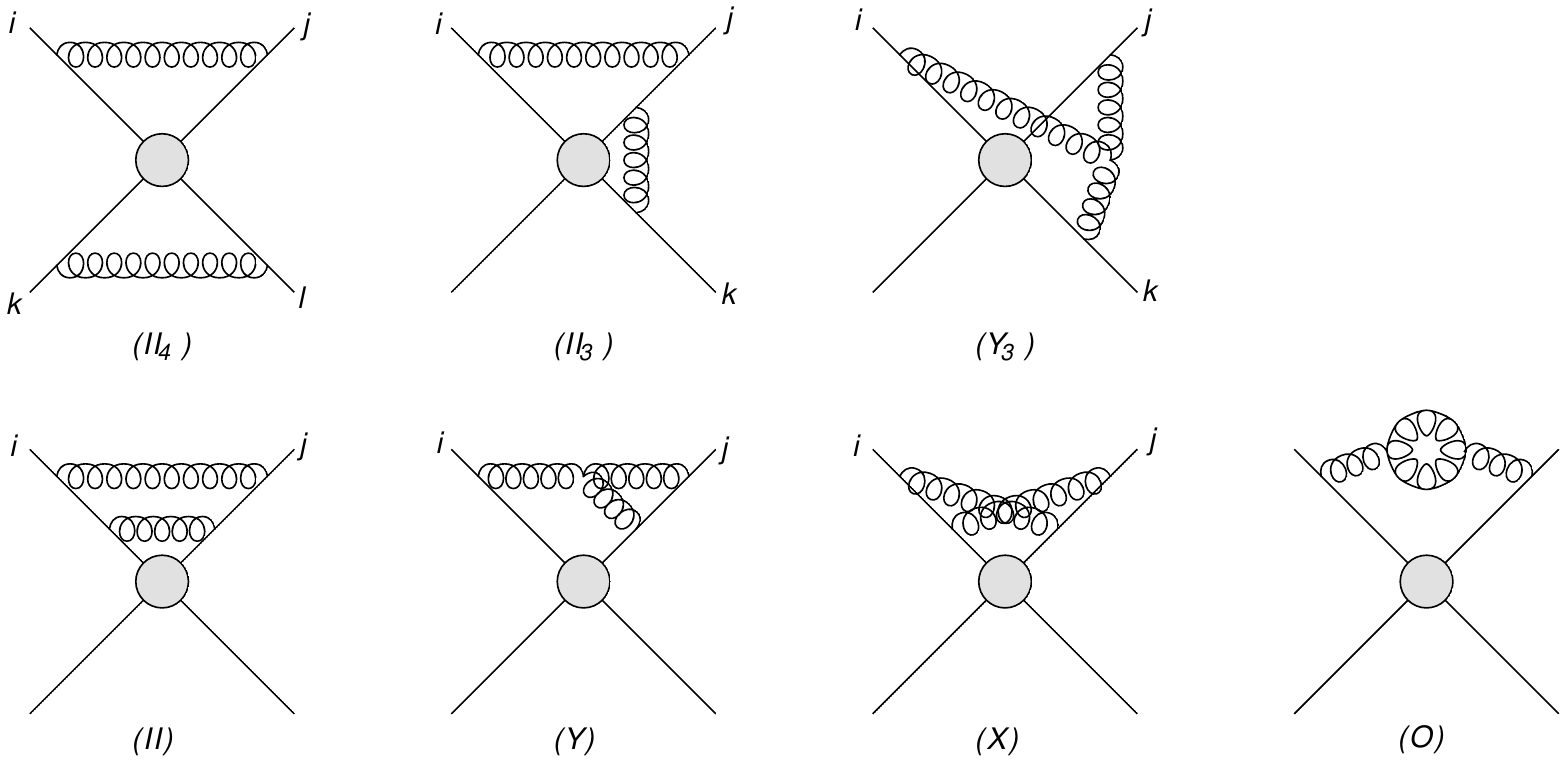}}
\caption{Two-loop soft topologies in $m$-point scattering, where only four
external lines are shown.}
\label{fig:2loopsoft}
\end{center}
\end{figure}

We consider tree-level $m$-point scattering dressed by soft gluons up to 
two-loop order. Possible soft topologies can then link two, three or four 
external lines. Examples are shown in figure~\ref{fig:2loopsoft}, where we 
label each topology, depicted in a schematic way (i.e. for brevity, we do not 
label momenta and their orientations)
We may also write the eikonal integrals that go with each soft topology, as
\begin{align}
{\cal I}^{(||_4)}_{ijkl}&=-\int\frac{d^Dk}{(2\pi)^D}\int\frac{d^Dl}{(2\pi)^D}
\frac{p_i\cdot p_j\,p_k\cdot p_l}{k^2\,l^2\,p_i\cdot k\,p_j\cdot k\,
p_k\cdot l\,p_l\cdot l},\notag\\
{\cal I}^{(||_3)}_{ijk}&=-\int\frac{d^Dk}{(2\pi)^D}\int\frac{d^Dl}{(2\pi)^D}
\frac{p_i\cdot p_j\,p_j\cdot p_k}{k^2\,l^2\,p_i\cdot k\,p_j\cdot k\,
p_j\cdot (k+l)\,p_k\cdot l},\notag\\
{\cal I}^{(Y_3)}_{ijk}&=-\int\frac{d^Dk}{(2\pi)^D}\int\frac{d^Dl}{(2\pi)^D}
\frac{V_{\mu\nu\rho}(k,l)\,p_i^\mu\,p_j^\nu\,p_k^\rho}{k^2\,l^2\,
(k+l)^2\,p_i\cdot k\,p_j\cdot l\,p_k\cdot (k+l)},\notag\\
{\cal I}^{(||)}_{ij}&=-\int\frac{d^Dk}{(2\pi)^D}\int\frac{d^Dl}{(2\pi)^D}
\frac{(p_i\cdot p_j)^2}{k^2\,l^2\,p_i\cdot k\,p_i\cdot (k+l)\,
p_j\cdot k\,p_k\cdot (k+l)},\notag\\
{\cal I}^{(Y)}_{ij}&=-\int\frac{d^Dk}{(2\pi)^D}\int\frac{d^Dl}{(2\pi)^D}
\frac{V_{\mu\nu\rho}(k,l)\,p_i^\mu\,p_j^\nu\,p_j^\rho}{k^2\,l^2\,
(k+l)^2\,p_i\cdot k\,p_j\cdot l\,p_j\cdot k},\notag\\
{\cal I}^{(X)}_{ij}&=-\int\frac{d^Dk}{(2\pi)^D}\int\frac{d^Dl}{(2\pi)^D}
\frac{(p_i\cdot p_j)^2}{k^2\,l^2\,p_i\cdot k\,p_i\cdot (k+l)\,
p_j\cdot l\,p_k\cdot (k+l)},\notag\\
{\cal I}^{(O)}_{ij}&=-\int\frac{d^Dk}{(2\pi)^D}\int\frac{d^Dl}{(2\pi)^D}
\frac{V_{\mu\nu\rho}(k,l)\,V_{\nu\sigma\rho}(k,l)\,p_i^\mu\,p_j^\sigma}
{(k^2)^2\,l^2\,(l-k)^2\,p_i\cdot k\,p_j\cdot k},
\label{2loopeik}
\end{align}
where 
\begin{equation}
V_{\mu\nu\rho}(k,l)\sim{\cal O}(k,l)
\label{Vscale}
\end{equation}
is the three-gluon vertex, which in this case couples together only soft 
momenta. Our notation for the eikonal integrals specifies which indices 
are coupled together by soft gluon emissions, and note that the ordering of
these indices can be important (e.g. ${\cal I}^{(Y)}_{ij}$ is not the same
as ${\cal I}^{(Y)}_{ji}$: the latter corresponds to a Y graph which is 
reflected with respect to the former). The infrared singular part of the 
two-loop gauge theory amplitude can now be written as
\begin{align}
{\cal A}^{(2)}_{\rm IR}&=-g_s^4\sum_{x\in\{s,t,u\}}
\left[\sum_{\langle ijkl\rangle}{\bf T}_i\cdot{\bf T}_j\,
{\bf T}_k\cdot{\bf T}_l\, 
{\cal I}_{ijkl}^{(||_4)}+\sum_{\langle ijk\rangle}\left(
{\bf T}_i\cdot{\bf T}_j\,{\bf T}_j\cdot{\bf T}_k\,{\cal I}^{(||_3)}_{ijk}
+\tilde{f}^{abc}\,{\bf T}_i^a\,{\bf T}_j^b\,{\bf T}_k^c\,
{\cal I}^{(Y_3)}_{ijk}\right)\right.\notag\\
&\left.\quad
+\sum_{\langle ij\rangle}\left(({\bf T}_i\cdot {\bf T}_j)^2\,
{\cal I}^{(||)}_{ij}
+\tilde{f}^{abc}\,{\bf T}_i^a\,{\bf T}_j^b\,{\bf T}_j^c\,{\cal I}^{(Y)}_{ij}
+{\bf T}_i^a\,{\bf T}_i^b\,{\bf T}_j^b\,{\bf T}_j^a\,
{\cal I}^{(X)}_{ij}+\tilde{f}^{abc}\,\tilde{f}^{bdc}\,{\bf T}_i^a\,{\bf T}_j^d
\,{\cal I}^{(O)}_{ij}\right)\right]\notag\\
&\quad\times\frac{n_x}{x},
\label{A2loop}
\end{align}
where the notation $\langle ij\ldots k\rangle$ denotes that one must sum over
all distinct multiples, and we have used the colour vertex factor of 
eq.~(\ref{colfac}). This expression is obtained by dressing the tree-level
interaction with all possible eikonal integrals, and is the two-loop 
generalisation of eq.~(\ref{A1IRdef}). Furthermore, we have taken a factor 
of $i^2=-1$ out of the eikonal integrals, so as to make manifest the phase
convention of eq.~(\ref{ampform}).\\

\begin{figure}
\begin{center}
\scalebox{1.0}{\includegraphics{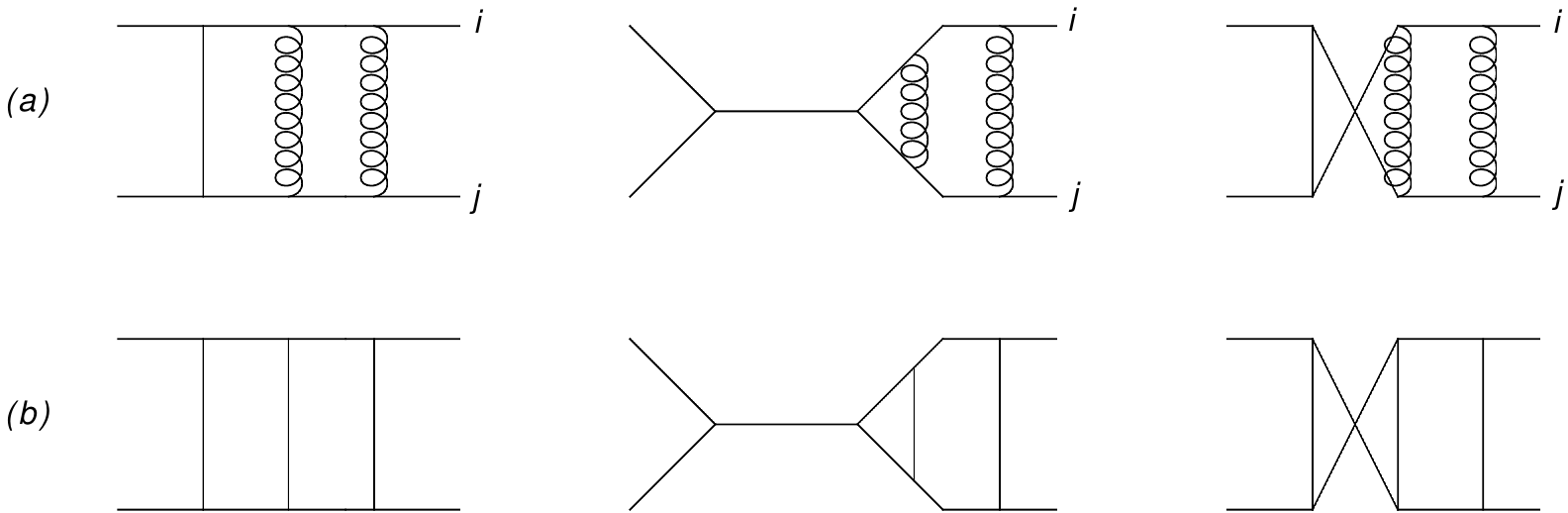}}
\caption{(a) Soft graphs corresponding to eq.~(\ref{I||ex}); (b) 
full QCD graphs in the corresponding BCJ relation.}
\label{fig:2loopex1}
\end{center}
\end{figure}

As in the one-loop case, we can match up terms in eq.~(\ref{A2loop}) into sets
of three, such that they correspond to the soft limit of a BCJ relation. There 
are two distinct classes of relation. Firstly, there are relations in which 
the same eikonal integral dresses different tree-level interaction graphs. This
is the only scenario that was possible at one-loop order, and the relevant 
three graphs are obtained by picking a particular term in eq.~(\ref{A2loop}), 
and keeping the sum over tree-level topologies $x$. As an example, consider 
the contribution
\begin{equation}
g_s^4\sum_x({\bf T}_i\cdot{\bf T}_j)^2\,{\cal I}^{(||)}_{ij},
\label{I||ex}
\end{equation}
obtained by selecting a particular pair in the first term in the second line
of eq.~(\ref{A2loop}). This corresponds to the graphs shown in 
figure~\ref{fig:2loopex1}(a). These indeed correspond to the soft limit of 
three full QCD graphs which enter a BCJ relation, namely those shown in 
figure~\ref{fig:2loopex1}(b). From eq.~(\ref{I||ex}), one may associate
a collected numerator and colour factor for each graph, according to
\begin{equation}
c^{(||)}_{x}=({\bf T}_i\cdot{\bf T}_j)^2\, c_x,\quad
\hat{n}^{(||)}_x=(p_i\cdot p_j)^2\,n_x(p_1,p_2,p_3),
\label{colnumex1}
\end{equation}
where the hat once again reminds us that such numerators are defined in the
soft limit, and may have arbitrary terms $\sim{\cal O}(k,l)$ added. The 
numerators thus defined satisfy the relations
\begin{equation}
\hat{n}^{(||)}_s-\hat{n}^{(||)}_u-\hat{n}^{(||)}_t=0,
\label{BCJ2loopex1}
\end{equation}
by virtue of the fact that this is satisfied by the tree-level numerators, 
and that these are all multiplied by a common factor. Although we have been
somewhat schematic in our notation, eq.~(\ref{BCJ2loopex1}) is indeed the
relevant limit of the BCJ relation with all basis momenta properly accounted
for, analagously to eq.~(\ref{nxijBCJ}).\\

It is clear that the above argument generalises to any set of three graphs
which have the same soft topology $Z$. That is, one may form soft numerators
\begin{equation}
\hat{n}^{(Z)}_x=\left.{\cal I}^{(Z)}_{ij\ldots k}\right|_{\rm num.}\,
n_x(p_1,p_2,p_3),
\label{numsoft}
\end{equation}
where $\left.{\cal I}^{(Z)}_{ij\ldots k}\right|_{\rm num.}$ denotes the 
numerator of the relevant eikonal integral, such that
\begin{equation}
\hat{n}^{(Z)}_s-\hat{n}^{(Z)}_u-\hat{n}^{(Z)}_t=0.
\label{BCJsoftgen}
\end{equation}
This then corresponds to the soft limit of a BCJ relation. \\

\begin{figure}
\begin{center}
\scalebox{0.8}{\includegraphics{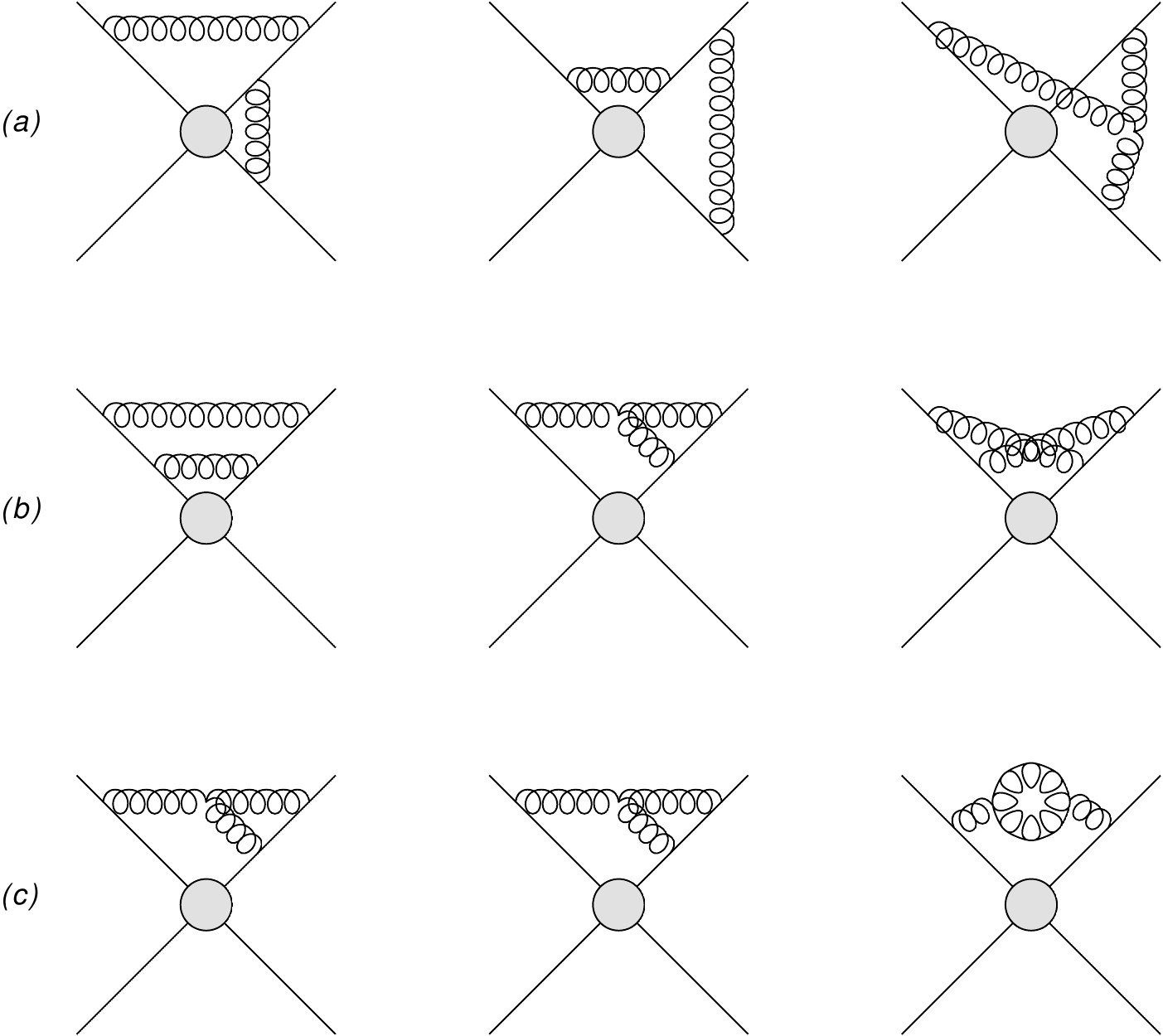}}
\caption{Sets of three graphs, each corresponding to the soft limit of a BCJ
relation, such that the hard interaction is the same in each term.}
\label{fig:BCJsoft}
\end{center}
\end{figure}

The second class of soft BCJ relations at two loop order is more complicated,
and consists of sets of three graphs in which the hard part of the interaction
(in this case a single tree-level topology) is the same in each term, but
the soft graphs are different. All possible examples are shown in 
figure~\ref{fig:BCJsoft}, and the full (non-soft) BCJ triples can be obtained
by replacing the soft gluon lines with solid lines, as in 
figure~\ref{fig:2loopex1}(a) and (b). 
Note that in figure~\ref{fig:BCJsoft}(c), the Y graph is related to itself,
analagously to the case of the triangles considered in 
section~\ref{sec:1loopBCJ}. For BCJ duality to be satisfied, there must
exist soft numerators satisfying
\begin{align}
\hat{n}^{(Y_3)}_{x,ijk}+\hat{n}^{(||_3)}_{x,kji}-\hat{n}^{(||_3)}_{x,ijk}&=0,
\notag\\
-\hat{n}^{(||)}_{x,ij}+\hat{n}^{(X)}_{x,ij}+\hat{n}^{(Y)}_{x,ij}&=0,\notag\\
\hat{n}^{(Y)}_{x,ij}+\hat{n}^{(Y)}_{x,ij}-\hat{n}^{(O)}_{x,ij}&=0,
\label{2loopBCJrels}
\end{align}
(where appropriate momentum arguments are understood) corresponding to each of 
the sets of graphs in figure~\ref{fig:BCJsoft}, and where 
$\hat{n}^{(Z)}_{x,i\ldots k}$ is the soft numerator associated with a diagram 
where soft graph $Z$ dresses tree-level topology $x$. As is clear from 
the eikonal integrals of eq.~(\ref{2loopeik}), the numerators of the graphs in 
the Feynman gauge do not automatically satisfy these relations. The numerators 
are given by
\begin{align}
\tilde{n}^{(||_3)}_{x,ijk}&=p_i\cdot p_j\, p_j\cdot p_k\,n_x,\notag\\
\tilde{n}^{(Y_3)}_{x,ijk}&=V_{\mu\nu\rho}(k,l)\,p_i^\mu\,p_j^\nu\,p_k^\rho\,n_x
\notag\\
\tilde{n}^{(Y)}_{x,ij}&=V_{\mu\nu\rho}(k,l)\,p_i^\mu\,p_j^\nu\,p_j^\rho\,n_x
\notag\\
\tilde{n}^{(||)}_{x,ij}&=(p_i\cdot p_j)^2\,n_x,\notag\\
\tilde{n}^{(X)}_{x,ij}&=(p_i\cdot p_j)^2\,n_x,\notag\\
\tilde{n}^{(O)}_{x,ij}&=V_{\mu\nu\rho}(k,l)\,V_{\nu\sigma\rho}(k,l)\,p_i^\mu\,
p_j^\sigma\,n_x,
\label{ntildedef}
\end{align}
where we have used a tilde to denote the fact that these soft numerators do not
respect BCJ duality. In order to find those that do, one must effect a 
generalised gauge transformation $\tilde{n}\rightarrow\hat{n}$. The general
form of such a transformation is given in eq.~(\ref{gengauge}), and consists
of redefining each numerator through superpositions of denominator factors. 
If, however, all we care about is verifying the infrared singularities of
gravity via the double copy, {\it we do not have to solve the BCJ relations
explicitly}. To see this, note the numerator factors depend on soft momenta
as
\begin{equation}
\tilde{n}^{(||_3)}_{x,ijk},\tilde{n}^{(||)}_{x,ij},\tilde{n}^{(X)}_{x,ij}\sim
{\cal O}(K^0),\quad \tilde{n}^{(Y_3)}_{x,ijk},\tilde{n}^{(Y)}_{x,ij}
\sim{\cal O}(K),\quad \tilde{n}^{(O)}_{x,ij}\sim{\cal O}(K^2),
\label{numscales}
\end{equation}
where $K\equiv(k,l)$ represents a soft momentum scale. Furthermore, all 
relevant denominator factors scale as at least ${\cal O}(K)$. Thus, from 
eq.~(\ref{gengauge}) one must have
\begin{align}
\hat{n}^{(||_3)}_{x,ijk}&=\tilde{n}^{(||_3)}_{x,ijk}+{\cal O}(K),\notag\\
\hat{n}^{(||)}_{x,ij}&=\tilde{n}^{(||)}_{x,ij}+{\cal O}(K),\notag\\
\hat{n}^{(X)}_{x,ij}&=\tilde{n}^{(X)}_{x,ij}+{\cal O}(K),
\label{numscales2}
\end{align}
and
\begin{equation}
\hat{n}^{(Y_3)}_{x,ijk},\hat{n}^{(Y)}_{x,ij},\hat{n}^{(O)}_{x,ij}
\sim{\cal O}(K).
\label{numscales3}
\end{equation}
We may summarise this more generally as follows. The BCJ dual numerators for
soft graphs involving multiple single gluon emissions between pairs of external
lines are the same as those in the Feynman gauge, up to corrections subleading
in soft momentum. These corrections will not lead to additional IR 
singularities, so can be ignored in the soft limit. This is consistent with
the first two BCJ relations in eq.~(\ref{2loopBCJrels}), which due to the 
subleading nature of $\hat{n}^{(Y_3)}_{x,ijk}$ and $\hat{n}^{(Y)}_{x,ij}$
amount to
\begin{equation}
\hat{n}^{(||_3)}_{x,ijk}=\hat{n}^{(||_3)}_{x,kji},\quad
\hat{n}^{(||)}_{x,ij}=\hat{n}^{(X)}_{x,ij},
\label{||eq}
\end{equation}
as is indeed observed already in eq.~(\ref{ntildedef}).\\

The BCJ dual numerators for graphs involving three gluon vertices off the 
eikonal lines are changed with respect to the Feynman gauge numerators, and 
are first order in soft momentum. Upon taking the double copy to gravity,
these numerators will be squared, whereas the denominators are unchanged.
Hence, by power counting, such graphs will not contribute infrared 
singularities in gravity. In other words, the very BCJ relations that one
has to invest effort in solving are irrelevant for the double copy! This 
also tells us that there are infrared singularities in QCD that vanish
when one takes the double copy to gravity. This ``information loss'' will, as
we will see, be a feature at higher loop orders. \\

The above remarks imply that we can take the double copy of 
eq.~(\ref{A2loop}) by keeping only the graphs with multiple dipole emissions.
Their numerators will be given by eq.~(\ref{numscales2}), where we can safely 
neglect the corrections which are subleading in soft momenta. Performing the
double copy procedure gives a 2-loop gravity amplitude~\footnote{For the 
overall phase factor, see the footnote on p. 12.}
\begin{align}
{\cal M}^{(2)}_{\rm IR}&=-i\left(\frac{\kappa}{2}\right)^4
\sum_{x\in\{s,t,u\}}
\left[\sum_{\langle ijkl\rangle}{\cal I}_{ijkl}^{(||_4),{\rm grav.}}
+\sum_{\langle ijk\rangle}{\cal I}^{(||_3),{\rm grav.}}_{ijk}
+\sum_{\langle ij\rangle}\left({\cal I}^{(||),{\rm grav.}}_{ij}
+{\cal I}^{(X),{\rm grav.}}_{ij}\right)\right]\frac{n_x\,n_x}{x},
\label{M2loop}
\end{align}
where
\begin{align}
{\cal I}^{(||_4),{\rm grav.}}_{ijkl}&=\int\frac{d^Dk}{(2\pi)^D}
\int\frac{d^Dl}{(2\pi)^D}\frac{4(p_i\cdot p_j)^2\,(p_k\cdot p_l)^2}{k^2\,l^2\,
p_i\cdot k\,p_j\cdot k\,p_k\cdot l\,p_l\cdot l},\notag\\
{\cal I}^{(||_3),{\rm grav.}}_{ijk}&=\int\frac{d^Dk}{(2\pi)^D}
\int\frac{d^Dl}{(2\pi)^D}\frac{4(p_i\cdot p_j)^2\,(p_j\cdot p_k)^2}
{k^2\,l^2\,p_i\cdot k\,p_j\cdot k\,p_j\cdot (k+l)\,p_k\cdot l},\notag\\
{\cal I}^{(||),{\rm grav.}}_{ij}&=\int\frac{d^Dk}{(2\pi)^D}
\int\frac{d^Dl}{(2\pi)^D}\frac{4(p_i\cdot p_j)^4}
{k^2\,l^2\,p_i\cdot k\,p_i\cdot (k+l)\,p_j\cdot k\,p_j\cdot (k+l)},\notag\\
{\cal I}^{(X),{\rm grav.}}_{ij}&=\int\frac{d^Dk}{(2\pi)^D}
\int\frac{d^Dl}{(2\pi)^D}\frac{4(p_i\cdot p_j)^4}
{k^2\,l^2\,p_i\cdot k\,p_i\cdot (k+l)\,p_j\cdot l\,p_j\cdot (k+l)}.\notag\\
\label{2loopeikgrav}
\end{align}
This indeed agrees with an explicit calculation using the known 
gravitational eikonal factors, obtained using the Feynman rules of 
eqs.~(\ref{eikrulegrav}) and~(\ref{prop}). We have thus shown that, at two 
loop order, the infrared singularities of GR are consistent with those of QCD 
via the double copy procedure. \\

We can write eq.~(\ref{2loopeikgrav}) in a more recognisable form as follows.
Firstly, one has
\begin{equation}
{\cal I}^{(||_4),{\rm grav.}}_{ijkl}={\cal I}^{\rm grav.}_{ij}
{\cal I}^{\rm grav.}_{kl},
\label{Iprod}
\end{equation}
where the right-hand side contains the product of two gravitational one-loop
eikonal factors from eq.~(\ref{Iijcalc}). By collecting terms in 
eq.~(\ref{2loopeikgrav}), we can write the entire right-hand side in terms
of one-loop integrals. In particular, one has
\begin{align}
{\cal I}^{(II_3),{\rm grav.}}_{ijk}+{\cal I}^{(II_3),{\rm grav.}}_{kji}
&=\int\frac{d^Dk}{(2\pi)^D}
\int\frac{d^Dl}{(2\pi)^D}\frac{4(p_i\cdot p_j)^2\,(p_j\cdot p_k)^2}
{k^2\,l^2\,p_i\cdot k\,p_k\cdot l\,p_j\cdot(k+l)}\left[\frac{1}{p_j\cdot k}
+\frac{1}{p_j\cdot l}\right]\notag\\
&=\int\frac{d^Dk}{(2\pi)^D}
\int\frac{d^Dl}{(2\pi)^D}\frac{4(p_i\cdot p_j)^2\,(p_j\cdot p_k)^2}
{k^2\,l^2\,p_i\cdot k\,p_k\cdot l\,p_j\cdot k\,p_j\cdot l},
\label{Isum}
\end{align}
where in the second line we have used the {\it eikonal identity}
\begin{equation}
\frac{1}{p_j\cdot(k+l)}\left[\frac{1}{p_j\cdot k}+\frac{1}{p_j\cdot l}\right]
=\frac{1}{p_j\cdot k\,p_j\cdot l}.
\label{eikid}
\end{equation}
One thus has
\begin{equation}
{\cal I}^{(II_3),{\rm grav.}}_{ijk}+{\cal I}^{(II_3),{\rm grav.}}_{kji}
={\cal I}^{\rm grav.}_{ij}{\cal I}^{\rm grav.}_{jk}.
\label{Iprod2}
\end{equation}
Also, one has
\begin{align}
{\cal I}^{(||),{\rm grav.}}_{ij}+{\cal I}^{(X),{\rm grav.}}_{ij}
&=\int\frac{d^Dk}{(2\pi)^D}\int\frac{d^Dl}{(2\pi)^D}\frac{4(p_i\cdot p_j)^4}
{k^2\,l^2\,p_i\cdot k\,p_i\cdot (k+l)\,p_j\cdot (k+l)}\left[\frac{1}
{p_j\cdot k}+\frac{1}{p_j\cdot l}\right]\notag\\
&=\int\frac{d^Dk}{(2\pi)^D}\int\frac{d^Dl}{(2\pi)^D}\frac{4(p_i\cdot p_j)^4}
{k^2\,l^2\,p_i\cdot k\,p_i\cdot (k+l)\,p_j\cdot k\,p_j\cdot l}\notag\\
&=\frac{1}{2}\int\frac{d^Dk}{(2\pi)^D}\int\frac{d^Dl}{(2\pi)^D}
\frac{4(p_i\cdot p_j)^4}{k^2\,l^2\,p_i\cdot (k+l)\,p_j\cdot k\,p_j\cdot l}
\left[\frac{1}{p_i\cdot k}+\frac{1}{p_i\cdot l}\right]\notag\\
&=\frac{1}{2}\int\frac{d^Dk}{(2\pi)^D}\int\frac{d^Dl}{(2\pi)^D}
\frac{4(p_i\cdot p_j)^4}{k^2\,l^2\,p_i\cdot k\,p_i\cdot l\,p_j\cdot k\,
p_j\cdot l}\notag\\
&=\frac{1}{2}\left({\cal I}^{\rm grav.}_{ij}\right)^2,
\label{Iprod3}
\end{align}
where we have again used the eikonal identity of eq.~(\ref{eikid}), as well as
relabelling $k\leftrightarrow l$. Putting things together, we may rewrite
eq.~(\ref{M2loop}) using the results of eqs.~(\ref{Iprod}, \ref{Iprod2}, 
\ref{Iprod3}) to give
\begin{align}
{\cal M}^{(2)}_{\rm IR}&=-i\left(\frac{\kappa}{2}\right)^4
\sum_{x\in\{s,t,u\}}
\left[\sum_{\langle ijkl\rangle}{\cal I}^{\rm grav.}_{ij}
{\cal I}^{\rm grav.}_{kl}
+\sum_{(ijk)}{\cal I}^{\rm grav.}_{ij}{\cal I}^{\rm grav.}_{jk}
+\frac{1}{2}\sum_{\langle ij\rangle}\left({\cal I}^{\rm grav.}_{ij}\right)^2
\right]\frac{n_x\,n_x}{x},
\label{M2loop2}
\end{align}
where the notation $(ijk)$ denotes summing over all triples, such that the 
ordering of $i$ and $k$ is unimportant. The total soft prefactor (contents of
the square bracket in eq.~(\ref{M2loop2}) is easily checked to be the second
order term in the expansion of
\begin{equation}
\exp\left[i\left(\frac{\kappa}{2}\right)^2\sum_{\langle ij\rangle}
{\cal I}^{\rm grav.}_{ij}\right],
\label{exp}
\end{equation}
as expected from the known exponentiation and one-loop exactness of 
gravitational infrared divergences~\cite{Weinberg:1965nx,Naculich:2011ry,
Akhoury:2011kq}. This completes our analysis of BCJ duality and the double
copy in the soft limit at two-loop order. In the next section, we generalise
our remarks to all loop orders.

\section{General remarks}
\label{sec:gen}

In the previous two sections, we have seen that the structure of infrared 
singularities in QCD matches on to that of GR after applying the double
copy procedure. At one loop, we were able to apply the double copy due to the
fact that the Feynman gauge numerators for soft graphs satisfied
the appropriate BCJ relations in the soft limit, up to irrelevant corrections
(suppressed by powers of soft momentum). At two loops, these relations
could be separated into two classes: (a) those involving graphs whose
soft topology was the same, but whose underlying hard tree-level topology 
was different; (b) those involving graphs sharing the same hard interaction,
but having different soft topologies. BCJ relations of class (a) were 
automatically satisfied, due to the fact that the underlying tree-level 
numerators satisfy BCJ duality. Relations of class (b) were more complicated.
Numerators for graphs with no three gluon vertices off the external lines
(which we refer to as {\it dipole-like graphs} in the following) could be taken
to be the same as the Feynman gauge results up to subleading corrections in
soft momenta, which are irrelevant from the point of view of infrared 
singularities. Numerators of graphs involving three-gluon vertices off the
external lines did not automatically satisfy BCJ relations, and thus would
have to be modified by generalised gauge transformations in order to write
down a BCJ-dual representation of a QCD amplitude in the soft limit. 
However, these graphs are at least linear in soft momenta, and thus vanish
upon taking the double copy to gravity. This allowed us to verify that the
IR singularities of gravity are correctly reproduced by double copying the
QCD results at two-loop order, without having to explicitly solve the BCJ
relations.\\

The aim of this section is to argue that this argument generalises to all
loop orders. This is possible because we have already seen all of the 
necessary ingredients at two loop order. Firstly, the fact that the infrared 
limit BCJ relations fall into the two classes given above is generally true, 
independent of the loop order~\footnote{At one loop, as we have seen, only 
class (a) occurs.}. Then BCJ relations 
of class (a), and involving a given soft topology $(Z)$, take the generic form
of eq.~(\ref{BCJsoftgen}), with numerators given by eq.~(\ref{numsoft}). As at 
two loop order, these relations are satisfied by virtue of the fact that the 
tree-level numerators satisfy the BCJ relation, and are multiplied by a common 
factor.\\

BCJ relations of class (b) are again more complicated, and can be split into
two further subclasses: (i) those involving two dipole-like graphs and a graph
containing at least one three gluon vertex off the external lines; (ii) 
those involving three graphs with at least one three gluon vertex off the 
external lines. By the same power counting arguments that were used in the 
previous section, the numerators of dipole-like graphs are the same as their
Feynman gauge counterparts up to corrections which are subleading in soft
momenta (i.e. which do not contribute infrared singularities). Two dipole 
graphs which enter the same BCJ relation are related by a permutation of two
gluon emissions on an external line (e.g. figure~\ref{fig:BCJsoft}(a))
~\footnote{In the language of~\cite{Gardi:2010rn,Gardi:2011wa,Gardi:2011yz},
such diagrams are in the same multiparton web.}. 
Thus, BCJ relations of subclass (i) set the numerators of such graphs to be
equal, which is indeed satisfied in the Feynman gauge, which associates with 
a single soft gluon emission between lines $i$ and $j$ a contribution 
$2(p_i\cdot p_j)$, independently of any other gluon emissions. BCJ relations
of subclass (ii) are not satisfied by the Feynman gauge numerators for the 
relevant graphs. However, these numerators are all at least ${\cal O}(K)$ 
(where $K$ is an arbitrary soft momentum), and remain so after performing a 
generalised gauge transformation in line with eq.~(\ref{gengauge}). Thus, 
these graphs do not give infrared singularities after performing the double
copy. One thus does not need to solve the BCJ relations explicitly in order
to generate the IR divergences of the gravity amplitude. From three-loop order 
in the Feynman gauge, one must also consider graphs involving the 
four-gluon vertex off the eikonal lines. This also gives rise to numerators 
which involve non-zero powers of soft momenta, after rewriting such graphs in 
terms of cubic topologies, according to the usual BCJ procedure. \\

Note that there are also BCJ relations which mix up hard and soft information.
We have already seen an example of this at one-loop order, namely the relations
of eq.~(\ref{BCJTTb}) which each involve the same triangle numerator evaluated
with different momenta. Such relations also arise at higher loop orders, and
as at one loop order, the soft limit provides insufficient information 
regarding whether these relations are satisfied. This is again irrelevant, 
however, for verifying the infrared singularities in gravity: the explicit 
solution of such relations requires higher order terms in soft momentum, which
do not contribute infrared singularities. \\

The above remarks allow us to generalise eq.~(\ref{M2loop2}) to an arbitrary 
loop order, as~\footnote{Here we have absorbed factors of $i$ into the eikonal
integrals, rather than show these explicitly as in eq.~(\ref{ampform}).}
\begin{align}
{\cal M}^{(n)}_{\rm IR}&=\left(\frac{\kappa}{2}\right)^{2n}
\sum_{x\in\{s,t,u\}}\left[\sum_{m=2}^{2n}\sum_{\langle i_1\ldots i_{m}\rangle}
\tilde{\cal I}^{{\rm grav.}}_{i_1\ldots i_m}\right]\frac{n_x\,n_x}{x},
\label{Mnloop}
\end{align}
where we again abbreviate $n_x\equiv n_x(p_1,p_2,p_3)$.
In this formula, $\tilde{\cal I}^{{\rm grav.}}_{i_1\ldots i_m}$ is the eikonal
integral factor due to the sum of all dipole emissions that link lines
$i_1,i_2\ldots i_m$. The second sum in the square brackets in 
eq.~(\ref{Mnloop}) is then over all multipoles $\langle i_1\ldots i_m\rangle$ 
that are linked by such dipole emissions. The first sum is over all 
possible numbers $m$ of external lines. As in the two-loop analysis of the 
previous section, one may collect terms in eq.~(\ref{Mnloop}) into products 
of one-loop eikonal integrals, via multiple applications of the (higher-order)
eikonal identity, and thus rewrite eq.~(\ref{Mnloop}) as~\cite{Weinberg:1965nx}
\begin{align}
{\cal M}^{(n)}_{\rm IR}&=\left(\frac{\kappa}{2}\right)^{2n}
\sum_{x\in\{s,t,u\}}\left[\frac{1}{n!}\left(\sum_{\langle ij\rangle}
i\int\frac{d^Dk}{(2\pi)^D}\frac{2p_i\cdot p_j}{k^2\,p_i\cdot k\,p_j\cdot k}
\right)^n\,\right]\frac{n_x\,n_x}{x}.
\label{Mnloop2}
\end{align}
The tree-level amplitude dressed by soft gravitons to all orders, and obtained
via the double copy, is then given by
\begin{equation}
{\cal M}_{\rm IR}=\sum_{n=0}^\infty {\cal M}^{(n)}_{\rm IR}
=\exp\left[\sum_{\langle ij\rangle}
i\left(\frac{\kappa}{2}\right)^2\int\frac{d^Dk}{(2\pi)^D}\frac{2p_i\cdot p_j}{k^2\,p_i\cdot k\,p_j\cdot k}
\right]\frac{n_x\,n_x}{x},
\label{Mnloop3}
\end{equation}
in agreement with the known all-order structure of IR divergences in 
GR~\cite{Weinberg:1965nx}. Note that, once we had reached eq.~(\ref{Mnloop}),
the final result was guaranteed: what mattered was that the double copy
procedure correctly reproduces the fact that only dipole-like graphs appear
in gravity. The sum over all such graphs automatically leads to the 
exponentiation of the one-loop corrections~\footnote{This is the same argument
that occurs in the exponentiation of one-loop soft corrections in 
QED~\cite{Grammer:1973db,Weinberg:1965nx}. In that theory, however, one-loop
exactness is broken due to the presence of fermion bubbles.}. \\

This completes our argument that the all-order structure of IR divergences in
gravity is consistent with IR singularities in QCD, via the application of the
double copy procedure. This can therefore be taken as all-order evidence for
the double copy conjecture, albeit in a particular limit. Some further comments
are in order. Firstly, the general argument at ${\cal O}(\alpha_S^n)$ exhibits 
a property already remarked upon at two loops, namely that many singularities 
cancel upon taking the double copy to gravity. This means that the gravity side
of the correspondence cannot be used to constrain singularities on the gauge 
theory side. For this reason, the question posed earlier in the paper regarding
whether the QCD dipole formula has a gravitational origin appears to have a 
negative answer: the very singularities which could occur as corrections to the
dipole formula {\it vanish} when we move to gravity. \\

Secondly, we have here focused on the case of pure Yang-Mills theory and
General Relativity. However, the argument of this paper can also in principle
be applied in supersymmetric gauge theories / supergravity. Here we remark that
reference~\cite{BoucherVeronneau:2011qv} obtained amplitudes in 
${\cal N}\geq 4$ supergravity using the double copy procedure applied to gauge
theory amplitudes in ${\cal N}=4$ Super-Yang-Mills theory coupled with 
$0\leq{\cal N}\leq 4$ Super-Yang-Mills theory. One check on this calculation
was the demonstration that infrared singularities, after the double copy,
were consistent with exponentiation on the gravity side up to two loop order.
The results of this paper generalise this to all loop orders, and 
non-supersymmetric theories. In supersymmetric theories, it may be possible
to extend our arguments beyond the pure soft limit, as non-trivial information
can in principle be obtained from infrared 
singularities~\cite{BoucherVeronneau:2011nm,Dunbar:2012aj}. \\

Throughout the paper, we have seemingly ignored the fact that the double copy
relates pure Yang-Mills theory to General Relativity coupled to a dilaton and
two-form. The additional particles may be present on the gravity side of the
double copy, and could in principle affect the soft behaviour. However, we 
have here considered only the leading infrared singularity at each order in
perturbation theory, obtained by dressing a tree-level hard interaction 
consisting of graviton scattering. This will not be affected by the 
presence of additional particles, which decouple in the soft limit due to
power counting arguments~\cite{DeWitt:1967uc} (see also~\cite{Saotome:2012vy} 
for a recent discussion of this point)~\footnote{The basic argument is that the
gravitational Feynman rules are such that the numerator for a graph involving
e.g. a scalar loop is at least ${\cal O}(K)$, where $K$ is a soft momentum. By 
a similar argument to that presented in section~\ref{sec:2loop}, generalised 
gauge transformations would not change this.}. One may obtain subleading 
infrared singularities in the gravitational amplitude in principle by coupling 
higher order hard interactions (including those involving additional particles)
to the gravitational soft function discussed throughout the present paper. 
Whether these subleading singularities match up between the gauge and gravity 
theories is then purely a statement about the hard function (due to one-loop 
exactness on the gravity side), and amounts to whether the double copy is 
satisfied in full QCD. 

\section{Conclusion}
\label{sec:conclude}

In this paper, we have examined the soft limits of QCD (strictly speaking, 
pure gluodynamics) and GR, and showed that infrared singular contributions to
amplitudes in both theories match up with each other upon using the double 
copy procedure of~\cite{Bern:2010ue,Bern:2010yg}. The structure of IR 
divergences in QCD scattering amplitudes is still subject to some uncertainty,
and our current state of knowledge can be expressed in terms of the dipole
formula of~\cite{Gardi:2009qi,Gardi:2009zv,Becher:2009cu,Becher:2009qa}, plus 
possible corrections~\cite{Dixon:2009ur,Bret:2011xm,DelDuca:2011ae,
Vernazza:2011aa,Ahrens:2012qz}. By contrast, the structure of IR singularities
is known exactly in GR~\cite{Weinberg:1965nx,Naculich:2011ry,White:2011yy,
Akhoury:2011kq}, where they exponentiate in terms of one-loop graphs only. \\

Being able to take the double copy relies on the gauge theory amplitudes
displaying manifest BCJ duality~\cite{Bern:2008qj}. At one-loop order, we saw
that amplitudes in the soft limit, as calculated in the Feynman gauge, 
satisfied BCJ duality up to irrelevant corrections. A similar story occured 
at two-loop order, where only dipole-like graphs had numerators which were 
unmodified (at leading order in soft momentum) by generalised gauge 
transformations. These are the only graphs that survive
upon taking the double copy, so that one does not need to solve the BCJ 
relations explicitly. We could thus use the double copy procedure to 
``predict'' the structure of IR singularities in GR, finding exact agreement
with the known results. Our arguments imply that many singularities vanish
upon taking the double copy. If this were not the case, one could have used
singularities in gravity to constrain possible corrections to the QCD dipole
formula, or even to provide an underlying gravitational explanation for this.
Nevertheless, the arguments presented in this paper constitute all-loop level
evidence for the validity of the double copy conjecture, which may be more
significant in supersymmetric contexts. \\

As mentioned above, we have only considered the case where the hard interaction
consists of tree-level scattering, and hence the leading infrared singularity 
at each order in perturbation theory on the gravity side of the double copy. 
In principle, this could contain higher loop orders, and include the double 
copy scalar and two-form. To show that the double copy is 
satisfied for the resulting subleading IR singularities then requires BCJ-dual 
numerators for the hard interaction, which is not possible to all orders 
without a full proof of BCJ duality in QCD. \\

There are a number of further questions that can be addressed. It may be 
possible, for example, to reinterpret our results using the manifestly
BCJ-dual effective Lagrangian of~\cite{Bern:2010yg}. It would also be 
interesting to examine the full implications of the present analysis in
supersymmetric contexts. Finally, a thorough investigation of the role of
BCJ duality in non-supersymmetric gauge theory away from the soft limit would
potentially provide new insights into QCD and / or gravity. An intermediate
step in this regard might be to extend the present analysis to next-to-eikonal
order, using the technology of~\cite{Laenen:2008gt,Laenen:2010uz,White:2011yy}.

\section{Acknowledgments}
CDW is supported by the UK Science and Technology Facilities Council (STFC). 
We thank Lorenzo Magnea for comments on the manuscript; also Claude Duhr,
Einan Gardi, Eric Laenen and David Miller for discussions, together with the 
other participants of the SUPA soft gluon workshop. We also gratefully 
acknowledge the help of the anonymous referee, particularly regarding the 
one-loop BCJ relations in QCD.

\bibliography{refs.bib}
\end{document}